\documentclass{aastex}          
\usepackage{spr-astr-addons}    

\begin{document}

\title{Long-term \emph{Fermi} observations of Mrk\,421: clues for different non-stationary processes}
\shorttitle{Long-Term \emph{FERMI} OBSERVATIONS OF MRK\,421} 

\author{B. Kapanadze\altaffilmark{1,2,3}, A. Gurchumelia\altaffilmark{2,4}, M. Aller\altaffilmark{5}}

\altaffiltext{1}{Ilia State University, Colokashvili Av. 3/5, 0162 Tbilisi, Georgia} 
\altaffiltext{2}{E. Kharadze National Astrophysical Observatory, Mt. Kanobili, 0803 Abastumani, Georgia}
\altaffiltext{3}{INAF, Osservatorio Astronomico di Brera, Via E. Bianchi 46, 23807 Merate, Italy} 
\altaffiltext{4}{I. Javakhishvili State University, Chavchavadze Av. 3, Tbilisi 0128, Republic of Georgia}
\altaffiltext{5}{Astronomy Department, University of Michigan, Ann Arbor, MI 48109-1107, USA}

\begin{abstract}
This paper presents the gamma-ray spectral and timing results from the long-term regular observations of Mrk\,421 with the Large Area Telescope (LAT) onboard \emph{Fermi} during 2008\,August--2023\,August. We discerned six periods of the relatively stronger 0.3--300\,GeV activity compared to other time intervals. The baseline brightness level varied on timescales from several months to years during these periods, which was superimposed by shorter-term flares of the different asymmetry. The latter are explained  by various interplay between the light-crossing, particle acceleration and cooling timescales. The source also  frequently exhibited   two-peak  flares, to be triggered by the propagation of forward and reverse shocks after collision between the "shells" of high-energy plasma, moving with different speeds down the jet. The strongest long-term flaring activity was recorded during 2012\,June--2013\,October and 2017\,October--2018\,March when the source was mostly brighter than 10$^{-7}$ph\,cm$^{-2}$s$^{-1}$ in the 0.3--300\,GeV energy range and robustly detectable even on intraday timescales. We detected 25 instances of intraday variability and a large number of the flux doubling/halving instances, allowing to constrain the upper limit to the  emission zone size to be in the range  of 1.3$\times$10$^{16}$\,cm--1.1$\times$10$^{18}$\,cm. The source generally showed a lognormal variability in the LAT energy range, explained as an imprinting of the disc nonstationary processes on the jet, proton-initiated hadronic cascades or random fluctuations in the particle acceleration rate. Most of the 0.3--300\,GeV spectra were well-fit with a simple power-law model and showed a very broad range of the photon-index from $\Gamma$$\sim$2.8 down to $\Gamma$$\sim$1.2, with the mean values $\Gamma_{\rm mean}$=1.75--1.84 and distribution peaks $\Gamma_{\rm p}$=1.73--1.82 during the periods of strong LAT-band activity. Our spectral study also revealed the features of inverse-Compton upscatter of X-ray photons in the Klein-Nishina regime, relativistic magnetic reconnection, first-order Fermi mechanism within the magnetic field of different confinement efficiencies and stochastic acceleration. 
\end{abstract}


\section{INTRODUCTION}
BL Lacertae objects (BLLs) represent a blazar subclass and exhibit extreme observational features (see, e.g., \citealt{b08}):  featureless spectra with two broadband SED peaks in the  in the $\nu F_{\rm \nu}$ representation, established by  relativistically-boosted non-thermal emission from the jet closely aligned with our line-of-sight (with viewing angles $\theta$$\lesssim$10\,deg and high bulk Lorentz factor $\Gamma
$$\sim$10; see, e.g., \citealt{f14}); flux variability across the entire electromagnetic range, with different strengths and amplitudes depending on the spectral range; strong and variable radio-to-X-ray polarization; strong $\gamma$-ray emission in the high-energy (HE, E$>$1\,MeV) and very-high-energy (VHE, E$>$100\,GeV) bands. These objects are sub-classified as low, intermediate and high-energy-peaked BLLs (LBL, IBL and HBLs, respectively), based on the radio-to-X-ray flux ratios \citep{p95,b01}. On the other hand, the sequence LBLs$\rightarrow$IBLs$\rightarrow$HBLs is characterized by the increasing synchrotron SED peak ($E_{\rm p}$) values, decreasing dominance of the gamma-ray flux over the lower-frequency emission and  bolometric luminosity \citep{b07}. Note that among the 67 BLLs to be TeV-detected so far, the vast majority are HBLs (84\%)\footnote{http://tevcat.uchicago.edu}. Therefore, these sources  should comprise highest-energy particles and the most violent acceleration processes \citep{ah09}. 

The lower-energy SED component is firmly explained as synchrotron radiation emitted by ultrarelativistic electrons (and, possibly, positrons) in the magnetized jet medium, owing to the absence of spectral lines and high polarization \citep{c08}. However, many problems remain to be solved related to the jet particle content, acceleration and unstable mechanisms. More problems persist with  unambiguous identification of the physical mechanisms responsible for the production of the higher-energy SED component, representing $\gamma$-rays in HBLS. Currently, two basic mechanisms are most frequently considered: in the leptonic models, ultra-relativistic electron/positron populations perform an inverse Compton (IC) upscatter of  their own low-energy synchrotron emission (synchrotron-self-Compton model, SSC; \citealt{m85} and references therein), or  the "seed" photons originating from outside the jet (external inverse Compton, EIC; see, e.g., \citealt{s94}): thermal emission from the accretion disc (AD), dusty torus, broad-line region (BLR), or from stellar clusters located near the jet emitting region \citep{c15}. However, HBLs do not show any significant features inherent to the external photon sources \citep{ah09,plot12}. One-zone SSC scenarios predict a correlated X-ray and VHE variability, especially during strong flares when the emission from a single region is expected to dominate the broadband SED \citep{ah09}. 

Alternatively, hadronic or lepto-hadronic scenarios incorporate specific gamma-ray emission mechanisms to solve the difficulties with the leptonic models \citep{b10}. Namely, the so-called synchrotron-proton blazar (SPB) model and its modified versions state that the significant portion of the jet kinetic or magnetic power is used to accelerate protons (along with electrons) in a strongly magnetized environment to the threshold of the photo-pion (p$\gamma$) production on the soft photon field, followed by various synchrotron-emitting pair cascades (see, e.g. \citealt{m93,ah00,c20}). The lower-energy SED component still is an electron-synchrotron emission (primary and secondary electrons from the hadronic cascades), while the ultrarelativistic hadron population produces gamma-rays via the synchrotron mechanism. Moreover, the $p\gamma$-interaction  can produce either $\pi^0$ or $\pi^\pm$ mesons. The charged pions subsequently decay into muons and muon neutrinos, whereas muons themselves also decay to produce electrons, positrons, and neutrinos.  Consequently, the $\gamma$-photons can be emitted from the $\pi^0$-decay process, or by electrons from the $\pi^\pm\to \mu^\pm\to e^\pm$ decay \citep{b13}. Generally, the proton-emission should contribute to the lower-energy part of the $\gamma$-ray SED component, while those from the muon and pion cascades are expected to have higher energies and form the third SED component after the "mutual" higher-energy hump \citep{c15}. Finally, the $p\gamma$-interaction may also result in the Bethe–Heitler pair production as $p+\gamma \rightarrow e^\pm$ \citep{sol22}.

Intense timing/spectral variability studies in different spectral bands and checking for the multi-wavelength (MWL) correlations allow us to discern the viable emission scenario and unstable physical processes operating in blazar jets. These studies are particularly important in the gamma-ray energy range, since this emission is associated with the highest-energy particles, which lose energy very quickly and exist only in the vicinity of the acceleration sites. 

Mrk\,421   is a nearby ($z$=0.031) HBL source and one of the brightest extragalactic X-ray/TeV objects, providing an unique X-ray space laboratory for solving the aforementioned problems. The \emph{Fermi}--LAT \citep{at09} is collecting high-level $\gamma$-ray data since 2008\,August\,5, thus making us capable to perform a detailed timing and spectral study of our target.  Mrk\,421 was initially included in the \emph{Fermi}-LAT bright gamma-ray source list (0FGL, \citealt{ab09a}) using the existing LAT observations. The 1.5-yr LAT data revealed a variability up to a factor $\sim$3 above 0.3\,GeV \citep{ab11a}. During the MWL campaign performed in 2009\,January--June, the source showed considerable HE variability uncorrelated with those in other energy ranges \citep{al15a}. \cite{al15b} used the LAT observations performed in 2010\,March for constructing a broadband SED and study MWL correlations. No significant HE variability was detected, contrary to the X-ray and VHE bands. \cite{hov15} adopted the 2012--2013 LAT data to study  the radio-$\gamma$-ray cross-correlations, which could be exist only for  a very specific choice of the model parameters.  \cite{abey17} did not find any significant variability from the daily-binned 0.1--30\,GeV light curve from the time interval 2014\,April\,28--May\,4, and the broadband SED showed the higher-energy peak to be below $\sim$100\,GeV. \cite{b16} concluded the LAT-band variability during 2013\,January--March to be insignificant.  \cite{carn17} reported a major HE outburst in 2012 and other noticeable flares in 2013 and 2014. \cite{ban19} found only a mild variability in the \emph{Fermi}--GeV energy range during the strong X-ray and VHE outburst in 2010\,February\,10--26, while \cite{sh12} reported an intra-day flux variability (IDV) at energies $>$200 MeV on February 17 (although the light curve clearly shows the variability detection below the commonly accepted 3$\sigma$ threshold).  \cite{acc21} did not find any strong LAT-band flaring episodes during 2016\,December--2017\,June.  The 0.1--300\,GeV variability in 2012\,December--2018\,April was correlated with that observed in the optical and radio energy ranges, although the latter was lagging the GeV-band light curve by 30--100 days \citep{ar21}. A similar result was obtained by \cite{acc21} for the data collected till 2016\,June. The LAT data from the period 2022\,April--June (in the epoch of the X-ray polarimetric observations with IXPE) showed a flux variability by a factor of $\sim$3 \citep{abe24}.

\begin{table*} \small  \centering  \begin{minipage}{150mm}
\caption{\label{frac}  Summary of the LAT   observations of Mrk\,421 during 2008\,August--2023\,August. For each time integration,  the maximum, minimum and mean fluxes (Cols. 2, 3 and 4, respectively) and fractional variability amplitude (in percents; Col.\,5) are provided.}   \centering 
  \begin{tabular}{cccccccccccccccccc}     \hline
  Band (units)& Maximum & Minimum & Mean & $F_{\rm var}$   \\
(1) & (2) & (3) & (4)& (5)  \\
 \hline
 LAT 0.3--300\,GeV (2\,weeks, 10$^{-8}$ph\,cm$^{-2}$s$^{-1}$)  & 30.70(1.95) & 1.91(0.57)&7.19(0.04) &40.3(0.6)  \\
LAT 0.3--300\,GeV (1\,week, 10$^{-8}$ph\,cm$^{-2}$s$^{-1}$)  & 32.47(2.33)  &1.92(033) & 6.94(0.04) &43.3(0.5)\\
LAT 0.3--300\,GeV (4\,d, 10$^{-8}$ph\,cm$^{-2}$s$^{-1}$)   & 37.74(4.86) &1.83(0.61) & 6.71(0.04) &45.6(0.5) \\
LAT 0.3--300\,GeV (3\,d, 10$^{-8}$ph\,cm$^{-2}$s$^{-1}$)  &40.09(6.85) & 2.00(0.64) &6.69(0.04)&  46.2(0.5)\\
LAT 0.3--300\,GeV (2\,d, 10$^{-8}$ph\,cm$^{-2}$s$^{-1}$)   &42.40(3.63)&2.27(0.87) & 7.11(0.04) &43.4(0.5) \\
LAT 0.3--300\,GeV (1\,d, 10$^{-8}$ph\,cm$^{-2}$s$^{-1}$)  &50.21(5.54)&3.17(0.76) & 9.12(0.05)& 37.2(0.5)\\
  \hline \end{tabular} \end{minipage} \end{table*}

Our past studies of Mrk\,421 were mainly focused on the detailed X-ray spectral and timing properties of the source in the epochs of the strong X-ray flaring activity and/or densely-sampled observations with the X-ray Telescope onboard the satellite \emph{Swift} (\emph{Swift}-XRT; \citealt{b05}) during 2005\,March--2024\,December \citep{k18a,k18b,k16,k17a,k20,k24}. Our basic findings were as follows:  (i) extreme X-ray flaring by a factor of 3--20 on timescales of a few days--weeks; exceptionally strong flares with CR$>$100\,cts\,s$^{-1}$ (corresponding to de-absorbed fluxes $F_{\rm
0.3-10 keV}\gtrsim$5$\times$10$^{-9}$\,erg\,cm$^{-2}$s$^{-1}$) which occurred in 2008\,June, 2010\,February, 2013\,April and 2018\,January--February. While the TeV-band and X-ray variabilities were mostly correlated (with some exclusions characterized by  \textquotedblleft orphan\textquotedblright ~X-ray or TeV-band flares), the source sometimes varied in a complex manner in the MeV--GeV and radio--UV energy ranges (indicating that the MWL emission could not always be generated in a single zone); (ii) flux variability in X-rays and $\gamma$-rays showed a lognormal character, possibly indicating that the flux variability to be an imprint of the accretion disk instabilities onto the jet; (iii) extreme X-ray IDV during the strongest flares: flux doubling/halving timescales of 1--7\,hr, brightness fluctuations by up to 20\% within within a few hundred seconds (possibly related to the small-scale turbulent areas containing the strongest magnetic fields); (iv)  the vast majority of the 0.3--10 keV spectra were consistent with the log-parabolic model, which showed relatively low spectral curvature and correlations between the different spectral parameters (predicted in the case of the first- and second-order Fermi processes). The position of the synchrotron SED peak $E_{\rm p}$ underwent an extreme variability on diverse timescales between the energies $<$0.1\,keV and $>$29\,keV, with a frequent occurrence of the hard X-ray-peaking spectra in higher states (rarely observed in BLLs). The photon index showed very hard values on some occasions, hinting at the possible presence of a jet hadronic component; (v) very fast transitions  of logparabolic-to-powerlaw spectra, most plausibly caused by turbulence-driven relativistic magnetic reconnection (RMR).

The aforementioned MWL campaigns were triggered in the epochs of enhanced X-ray and TeV-band activity, and, consequently, these studies are biased towards the high states of the source. On the contrary, the LAT-band timing and spectral properties in the intermediate and low $\gamma$-ray states have been poorly presented in the literature. Since the spectral properties of Mrk\,421 in the MeV--GeV energy range was studied by a few authors in the restricted time span, no statistical treatment of the spectral properties and the corresponding physical implications are provided. By expanding our study on the 15-yr period of the LAT operations (2008\,August--2023\,August), we have investigated the timing and spectral properties of Mrk\,421 on various timescales. Our results are based only on the robust detections of the source with LAT, contrary to some previous studies which also included the data from the 100--300\,MeV band (not recommended for HBL sources). Our  experimental results have been compared with those obtained in the framework of the recent theoretical studies and simulations, allowing us to draw conclusions about the unstable physical processes operating in the target's jet. For this purpose, we also checked for the interplay between the 0.3--300\,GeV flux variability and those observed in other spectral ranges with different instruments: the First G-APD Cherenkov Telescope (FACT; \citealt{a13}); \emph{Swift}-XRT, the Ultraviolet-Optical Telescope (UVOT; \citealt{r05}) and  the Burst Alert Telescope (BAT; \citealt{ba05}) onboard \emph{Swift}, MAXI \citep{m09}, different ground-based optical and radio telescopes.

The paper is organized as follows: Section\,2 encompasses the description of the data processing and analyzing procedures. Section\,3 is devoted to the results of the 0.3--300\,GeV timing study and the contemporaneous MWL variability of Mrk\,421. The MeV--GeV spectral properties and variability are provided in Section\,4. We discuss physical implications from our experimental results and draw the corresponding conclusions in  Section\,5. Finally, the summary of our study is given in Section\,6.

\section{DATA REDUCTION and ANALYSIS }

The LAT data of Mrk\,421 were retrieved from the \emph{Fermi} data server\footnote{https://fermi.gsfc.nasa.gov/ssc/data/access/} and processed  by using the software \texttt{Science Tools} (version 2.2.0, provided by the Fermi-LAT collaboration\footnote{http://fermi.gsfc.nasa.gov/}). We followed the standard procedure provided by the Fermi Science
Support Center\footnote{https://fermi.gsfc.nasa.gov/ssc/data/analysis/} (FSSC). Namely,  the instrument response function \texttt{P8R3\_SOURCE\_V2} and the maximum likelihood method \texttt{GTLIKE}\footnote{http://fermi.gsfc.nasa.gov/ssc/data/analysis/documentation/
Cicerone/Cicerone\_Likelihood} were adopted. In order to extract the photon flux and spectral information, we selected the 0.3--300\,GeV energy range due to the following reasons: (i)  the LAT effective area is larger ($>$0.5m$^2$); (ii) the angular resolution is relatively good in this energy range (the 68\% containment angle smaller than 2$\textdegree$ versus 3.5$\textdegree$ at 100\,MeV); (iii) minimize contamination from misclassified cosmic rays \citep{at09}. Therefore, there are smaller systematic errors and the spectral fit is less sensitive to possible contamination from unaccounted, transient neighbouring sources \citep{ab11a}.

\begin{figure*}[ht!]
  \includegraphics[trim=6.0cm 0.65cm 8cm 0cm, clip=true, scale=0.88]{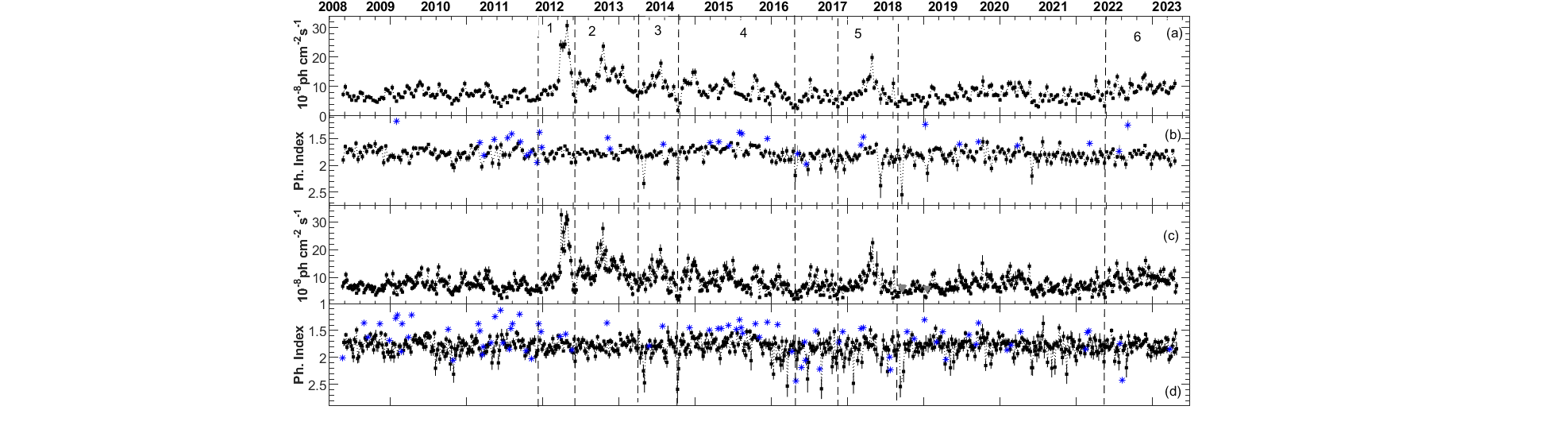}
  \includegraphics[trim=6.0cm 0cm 8cm 0cm, clip=true, scale=0.88]{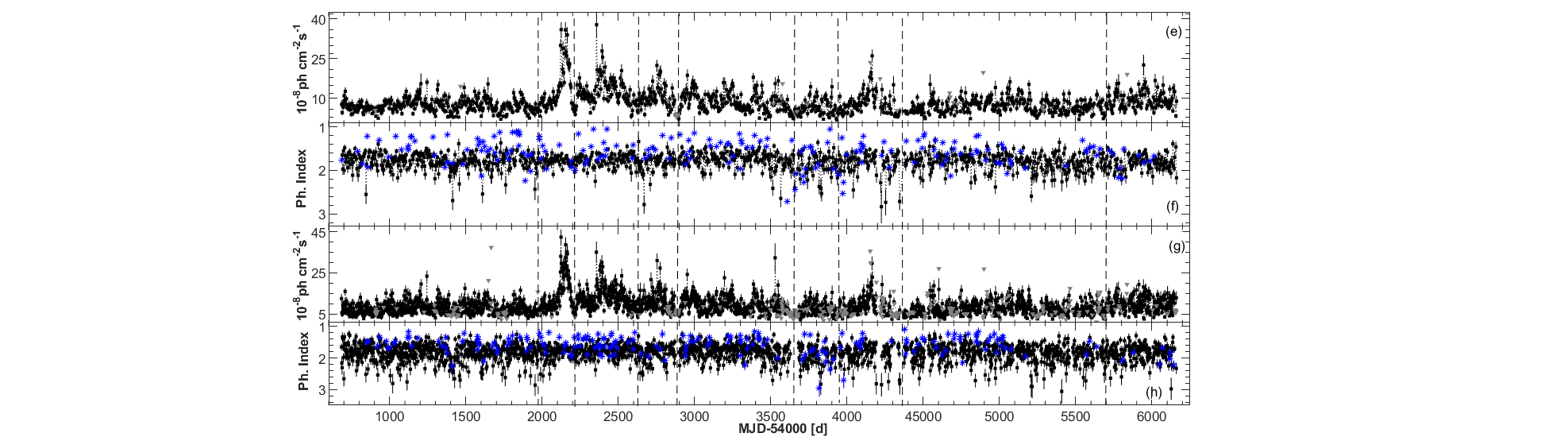}
\vspace{-0.3cm}
 \caption{\label{hist} 0.3--300\,GeV photon flux and  indices plotted versus time by using various time bins: panels (a)--(b): 2 weeks; (c)--(d): 1 week; (e)--(f): 4 days; (g)--(h): 2 days. The power-law and logparabolic photon indices are plotted with black points and blue asterisks, respectively. The downward gray triangles show the upper limits to the 0.3–300 GeV flux when $TS$$<$9 and/or $N_{\rm pred}$$<$8. The vertical dashed line denotes a boundary between the different time intervals listed in Table\,\ref{periods}. }   \end{figure*}

\begin{table*}  \begin{minipage}{170mm} 
  \caption{\label{periods} Summary of the LAT-band of Mrk\,421 in different periods. The maximum, minimum and mean LAT-band flux values (Columns 4, 5 and 6, respectively) are derived from one-weekly integrated LAT data and given in units of 10$^{-8}$ph\,cm$^{-2}$s$^{-1}$; fractional variability amplitude (Col. 7) -- in per cent. }    \centering
  \begin{tabular}{cccccccccccc}  \hline
 Per. & Dates (UTC) &MJDs& $F_{\rm max}$  &  $F_{\rm min}$ & $F_{\rm mean}$ & $F_{\rm var}$   \\
 (1)	& (2) &	(3)&	(4) &	(5)&	(6)&	(7)\\
    \hline
1&2012-03-01--10-15&55988--56216&	32.47(2.33)&4.62(0.70)&	14.04(0.23)	&60.4(1.8)  \\
2&2012-10-17--2013-10-25&(56)216--589&27.54(2.24)&3.93(0.88) &12.89(0.18)	&51.7(1.5)  \\
3&	2013-10-26--2014-08-19&	(56)590--888&20.03(1.83)&	2.25(0.80)&	9.40(0.19)&36.3(2.1)  \\
4&2014-08—20--2016-09-18&56889--57649&	16.79(1.70)&	1.91(0.69)&	8.46(0.10)	&34.8(1.6) \\
5&2017-07-03--2018-07-25&57938--58325&22.41(1.86)&	2.20(0.63)&	7.85(0.16)	&47.9(2.2)  \\
6&2022-06-01--2023-08-01&59696--60157&16.01(1.58)&	2.80(0.72)&	9.50(0.14)&	33.6(1.7) \\
  \hline \end{tabular} \end{minipage} \end{table*}

The events of the \textquotedblleft diffuse\textquotedblright ~class (\texttt{evclass}=128, \texttt{evtype}=3) (i.e, those with the highest probability of being photons)  from a region of interest (ROI) with the 10-deg radius centered at the location of Mrk\,421 were included in our analysis. The data were filtered by using the \texttt{gtselect} and \texttt{gtmktime} tools included in the aforementioned software: (a) events at zenith angles $>$90\,deg were discarded to avoid a contamination from the Earth-albedo photons, generated by cosmic rays interacting with the upper atmosphere; (b) photons recorded when the spacecraft's rocking angle was larger than 52\,deg were removed that greatly reduced  the contamination from Earth-limb photons; (c) time intervals with poor data quality, flagged as anything other than \textquotedblleft Good\textquotedblright ~ were also excluded. A background subtraction was performed my means of the XML model file, created by using the Python application \texttt{make4FGLxml.py} and incorporating: (i) the Galactic diffuse-emission component; (ii) the isotropic component, which is the sum of the extragalactic diffuse emission and the residual charged particle background; (iii) all $\gamma$-ray sources from the 4FGL catalogue within the 20\,deg radius from Mrk\,421. For the spectral modelling of our target in the entire 0.3--300\,GeV energy range, we adopted the log-parabola model \citep{m04}, similar to the catalogue:
\begin{equation} dN/dE=K(E/E_{\rm 0})^{-[\alpha+\beta log(E/E_{\rm 0})]},
 \vspace{-0.1cm} \end{equation}
with $E_{\rm 0}$ the reference energy; $\alpha$, the photon index at the energy $E_{\rm 0}$; $\beta$, measures the curvature around the peak; $K$, the normalization factor. However, most of the spectra obtained via the different time integrations (from two weeks down to intraday timescales) did not show a significant curvature (detection below the 2$\sigma$ significance). In such a situation, we re-launched the \texttt{GTLIKE} tool by adopting a simple power-law model $F(E)=KE^{-\Gamma}$, with $\Gamma$, the photon index throughout the entire energy range, and the 0.3--300\,GeV photon flux was derived along with the $\Gamma$-value.

\begin{figure*}[ht!] 
  \includegraphics[trim=5.9cm 1.0cm 0cm 0cm, clip=true, scale=0.90]{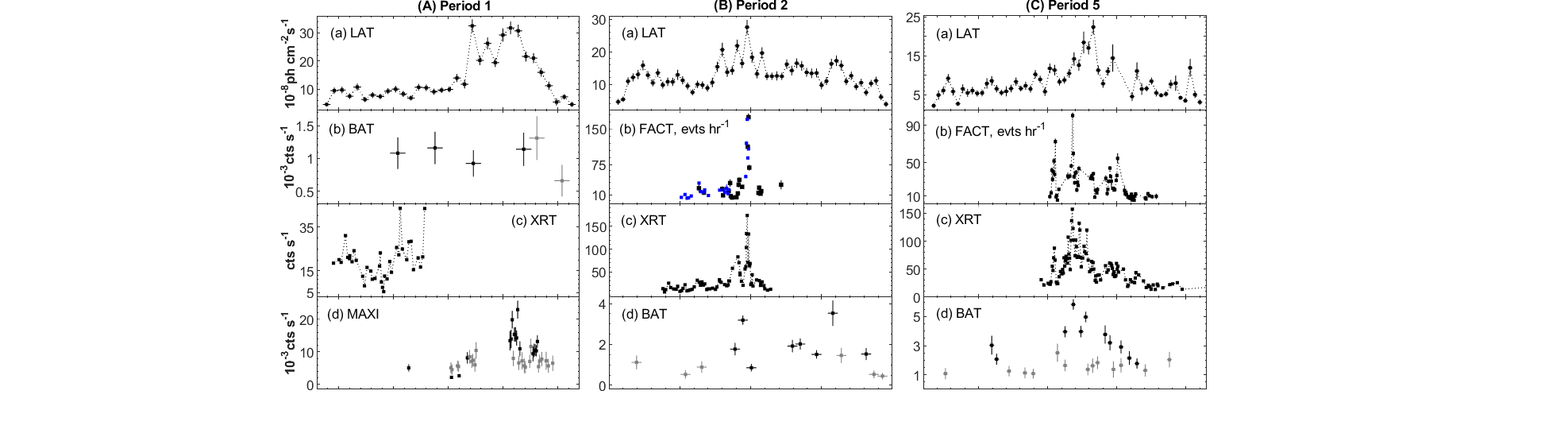}
\includegraphics[trim=5.9cm 1.1cm 0cm 0cm, clip=true, scale=0.90]{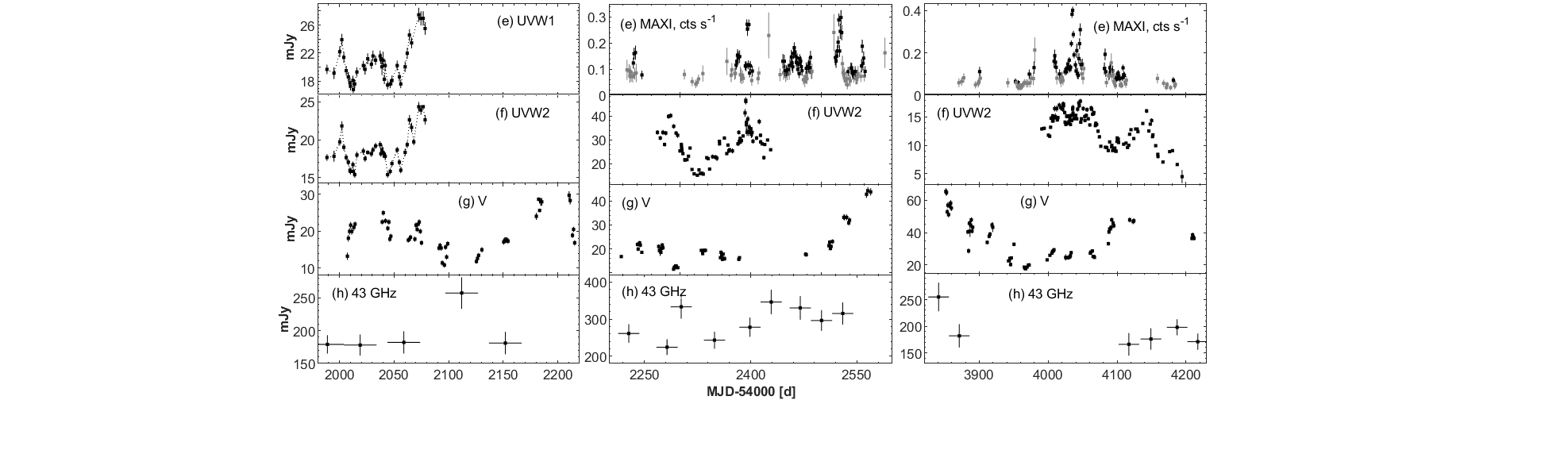} \vspace{-0.6cm}
 \caption{\label{per}  MWL light curves  in the periods listed in Table\,\ref{periods}.  While the XRT, MAXI, FACT, UVOT,  \emph{V}-band and UMRAO data (next page) data are plotted by using the 1-d integration, the BAT and LAT light curves are based on data binned every one week. The 43 GHz data are obtained approximately monthly. In the panels with the BAT and MAXI-band light curves, the black  and grey data points correspond to the detections of Mrk\,421 with 5$\sigma$ and (3--4)$\sigma$ significances, respectively.  In Period\,2, the blue points in the second panel from the top correspond to the VHE flux values from the observations with the different Cherenkov-type telescopes (originally provided in the different units and multiplied by the corresponding numbers to  make compatible with the FACT results presented in events\,hr$^{-1}$). }
 \end{figure*}

The spectral parameters of the sources within the ROI were left free during the minimization process, while those outside of this range were held fixed to the 4FGL catalog values (according to the common practice; see, e.g., \citealt{acc21}).  In order to reduce systematic uncertainties in the analysis, the photon index of the isotropic component and the normalization of both components were fixed (see \citealt{ab11b}). The light curves were generated using the time bins of different duration, repeating the likelihood analysis for each interval. In each case, the photon flux, photon index, curvature parameter, test-statistics (TS) and the number of the model-predicted photons $N_{\rm pred}$ of the source were determined. The target's detection significance is $\sqrt{TS}\sigma$ \citep{ab09b}. In order to have the target's robust detection, we employed the 3$\sigma$ threshold. Moreover, when the  number of the model-predicted counts $N_{\rm pred}$$\lesssim$8, such a detection is not robust. For example, even a small change in the time-bin width can result in significantly different values of the photon flux and spectral parameters. In such cases, we calculated the upper limit to the photon flux\footnote{According to the recipe provided on fermi.gsfc.nasa.gov/ssc/data/analysis/scitools/upper\_limits.html}.

\addtocounter{figure}{-1} \begin{figure*}[ht!] 
  \includegraphics[trim=5.9cm 0.85cm 0cm 0cm, clip=true, scale=0.90]{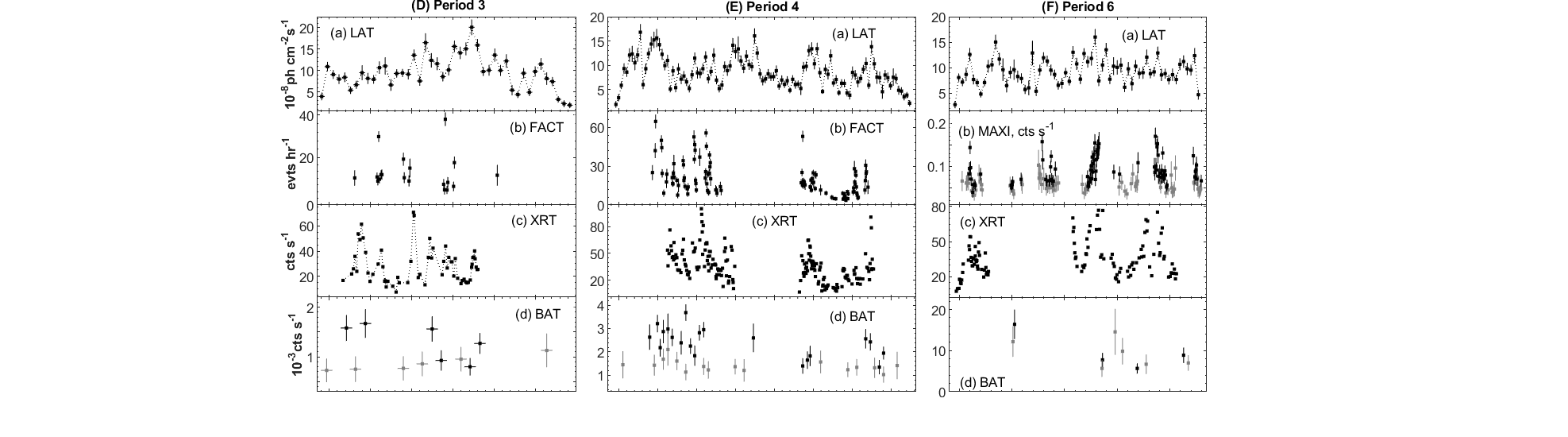}
\includegraphics[trim=5.9cm 1.1cm 0cm 0cm, clip=true, scale=0.90]{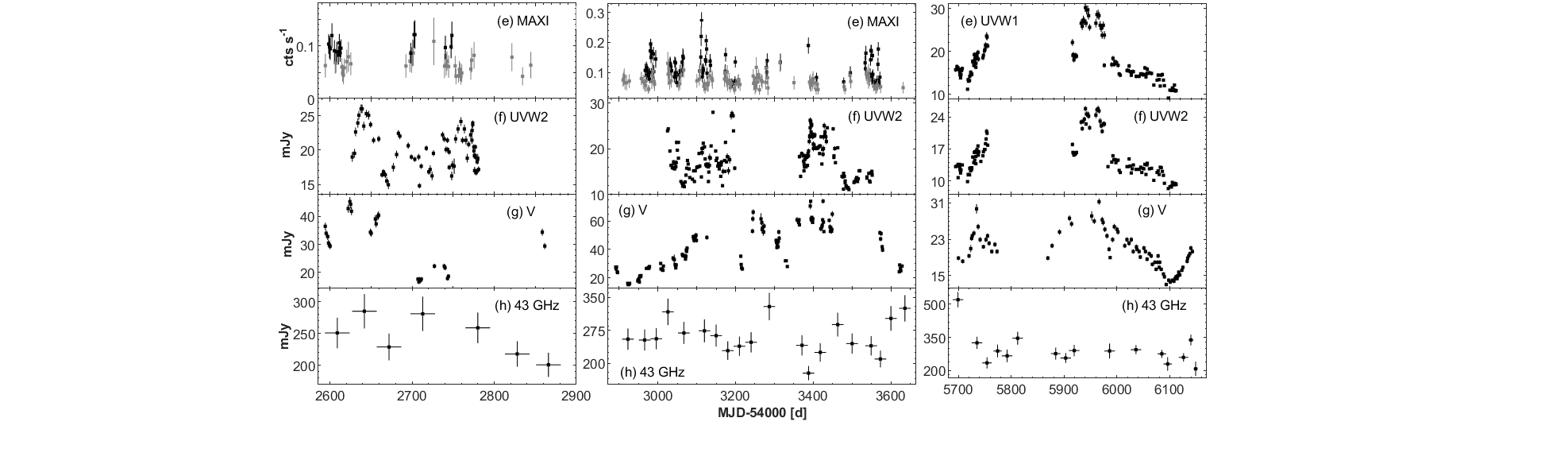}
\vspace{-0.7cm}  \caption{ - Continued.}  \end{figure*}

The source was monitored by the FACT telescope (Observatorio del Roque de los Muchachos, La Palma, Spain; \citealt{a13}) in the VHE range during the period of our study till the end of 2021\,June, followed by the period of seasonal "invisibility", volcano eruption at La Palma and its aftermaths, as well as by some technical problems\footnote{http://www.fact-project.org}. 
The FACT collaboration published the results of a quick-look analysis promptly after each observation\footnote{See http://www.fact-project.org/monitoring}.  Generally, the background-subtracted  VHE excess rates are not corrected for the effect of changing energy threshold with various zenith distances and ambient light \citep{d15}.  However, we restricted our study only to the nights with a signal detected with a minimum significance of 3$\sigma$. In this case, more than 98\% of the FACT data are obtained with zenith distances small enough to not significantly influence the energy threshold of the analysis. Nevertheless, more than 84\% of these data are taken under light conditions which do not increase the analysis threshold (see, e.g., \citealt{k20}). 

We also collected  the publicly available MWL  data  from the sources as follows: (a) the long-term Whiplle data included on the VERITAS website\footnote{http://veritas.sao.arizona.edu/veritas-science/mrk-421-long-term-lightcurve} (originally published by \cite{acc14}; (b) the X-ray, UV, optical and radio data from our past studies \citep{k16,k17a,k18a,k18b,k20,k24}, the AAVSO International Database\footnote{https://www.aavso.org/aavso-international-database}, University of Michigan Radio Astronomy Observatory (UMRAO) database\footnote{https://dept.astro.lsa.umich.edu/datasets/umrao.php} and  Large VLBA Project BEAM-ME program\footnote{https://www.bu.edu/blazars/BEAM-ME.html}; (c) the background-subtracted 2--20\,keV data obtained with \emph{MAXI}, which are publicly available on the mission's website\footnote{http://maxi.riken.jp}. As  customary for the coded-mask devices,  the retrieved, 1-d binned data was filtered by adopting the 5$\sigma$ detection threshold.  However, we also plotted the 2--20\,keV fluxes corresponding to the detections with (3--4)$\sigma$ significances for discerning the time intervals of relatively enhanced hard X-ray activity of the target. Furthermore, we used the 15--150\,keV \emph{Swift}-BAT data retrieved from the website of the Hard X-ray Transient Monitor program\footnote{http://swift.gsfc.nasa.gov/results/transients/weak/Mrk421} \citep{kr13}. Since the target was relatively rarely detectable with 5$\sigma$ significance (the BAT also to be a coded-mask device; see, e.g., \citealt{ba05}), we re-binned the orbit-resolved BAT data  via the \texttt{HEASOFT} task \texttt{REBINGAUSSLC}, using the 1-week integration.

\section{The LAT-band flux Variability on Various Timescales}

\subsection{Overall 0.3--300\,GeV variability}

Generally, Mrk\,421 is the brightest LAT-band source among the HBL objects and detectable even on intraday timescales during the strong HE $\gamma$-ray flares. The  LAT-band light curves constructed by using the different time integrations (from 2 weeks down to 2\,d) are presented in Figure\,\ref{hist}. In the period of our study, the source was not detectable with the LAT from the two-week binned 0.3--300\,GeV data only once: it was not observed during 2018\,March\,20--April\,9 [MJD\,(58)197--217] and, therefore, no good time intervals (GTIs) were available. This time interval corresponded to three bins in the case of the one-weekly integrated data, and there were another three occasions corresponding to the target's detection below the  3$\sigma$ threshold from the 1-week-binned data, after adopting the filtering criteria described in Section\,2.1. In the case of the shorter integration times, Mrk\,421 was  detected with a significance of at least 3$\sigma$ and $N_{\rm pred}$$\gtrsim$8 for 95, 90, 86 and 51 per cents of the 4, 3, 2 and 1-day bins, respectively.   A summary of the corresponding light curves is provided  in Table\,\ref{frac}, presenting the maximum, minimum and  mean values of each data train. Using the standard $\chi^2$-test, we define a source as variable if the probability that its flux is constant is less than 10$^{-3}$. The last column presents  the fractional variability amplitude and its error as follows \citep{v03}:
\begin{equation} 
 F_{\rm var}=(S^2-\overline{\sigma^2_{\rm err}})^{1/2} /{\overline{F}} \\
 \end{equation}
with  $S^2$, the sample variance; $\overline{\sigma^2_{\rm err}}$, the mean square error; $\overline{F}$, the mean flux. We see that the 0.3--300\,GeV photon flux frequently was higher than the level of 10$^{-7}$ph\,cm$^{-2}$s$^{-1}$. Note that such states has been very rarely observed  for other HBLs.  

The highest historical 0.3--30\,GeV flux of about 10$^{-6}$ph\,cm$^{-2}$s$^{-1}$ was recorded on 2013\,April\,15, during the two subsequent 1-hr segments [MJD\,56397.(33--42)]. However, the source was not detected securely above this threshold, taking into account the associated uncertainties. Note that Mrk\,421 was detectable during the even shorter, 0.6-hr time interval on 2012\,July\,16 [MJD\,(56124.)47–50] coinciding with the strongest MeV--GeV band activity of Mrk\,421 since the start of \emph{Fermi} operations (see below).

\begin{figure*}[ht!] 
\includegraphics[trim=6.0cm 5.0cm -1cm 0cm, clip=true, scale=0.89]{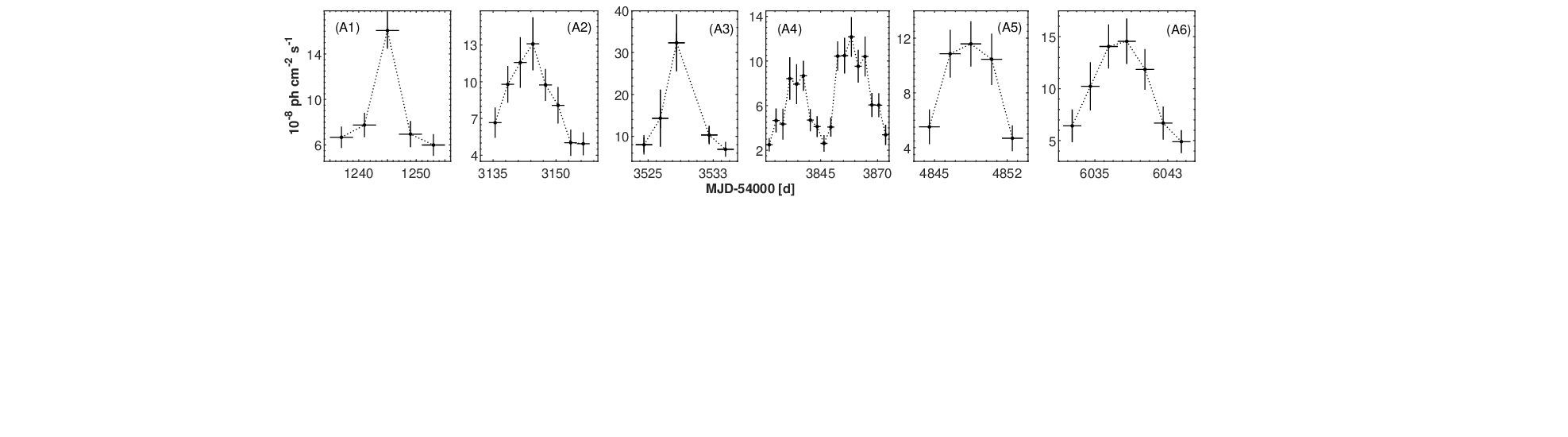}
\includegraphics[trim=6.0cm 4.7cm -1cm 0cm, clip=true, scale=0.89]{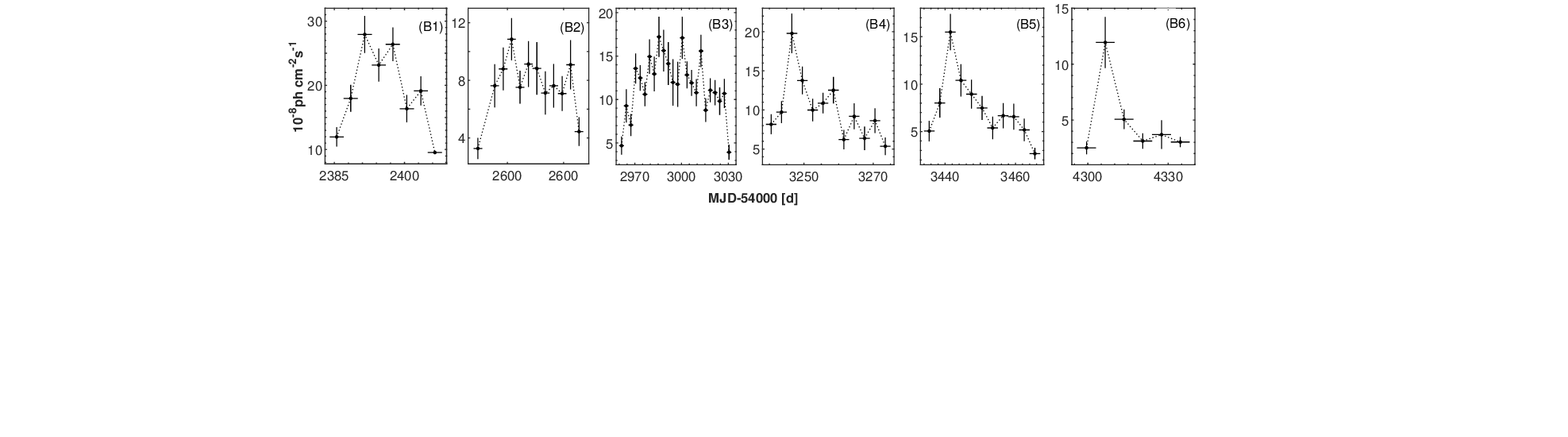}
\includegraphics[trim=6.0cm 4.8cm -1cm 0cm, clip=true, scale=0.89]{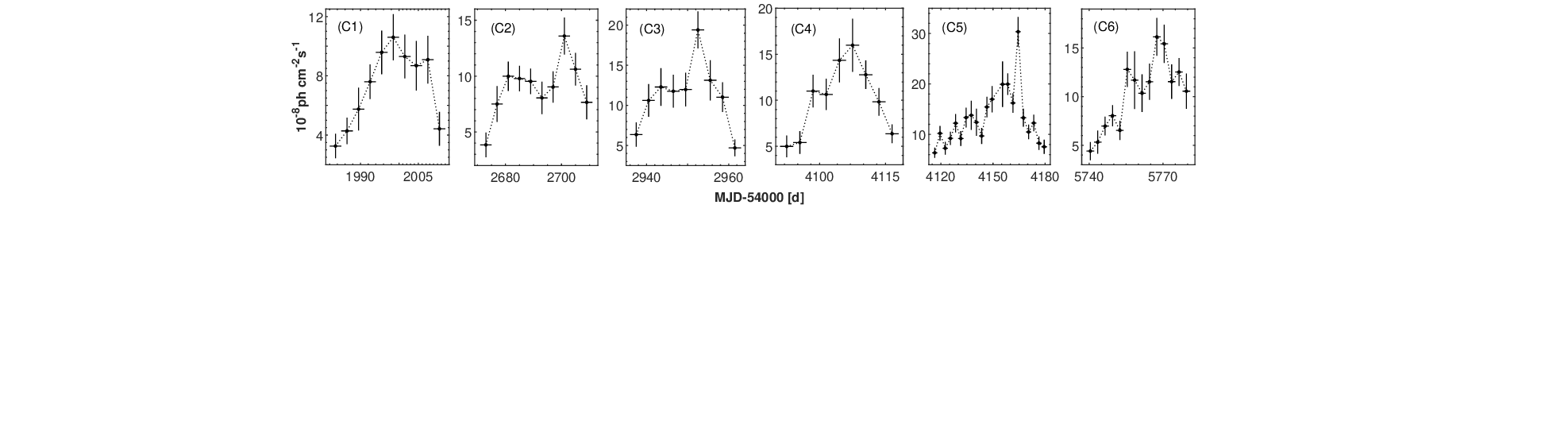}
\includegraphics[trim=6.0cm 4.8cm -1cm 0cm, clip=true, scale=0.89]{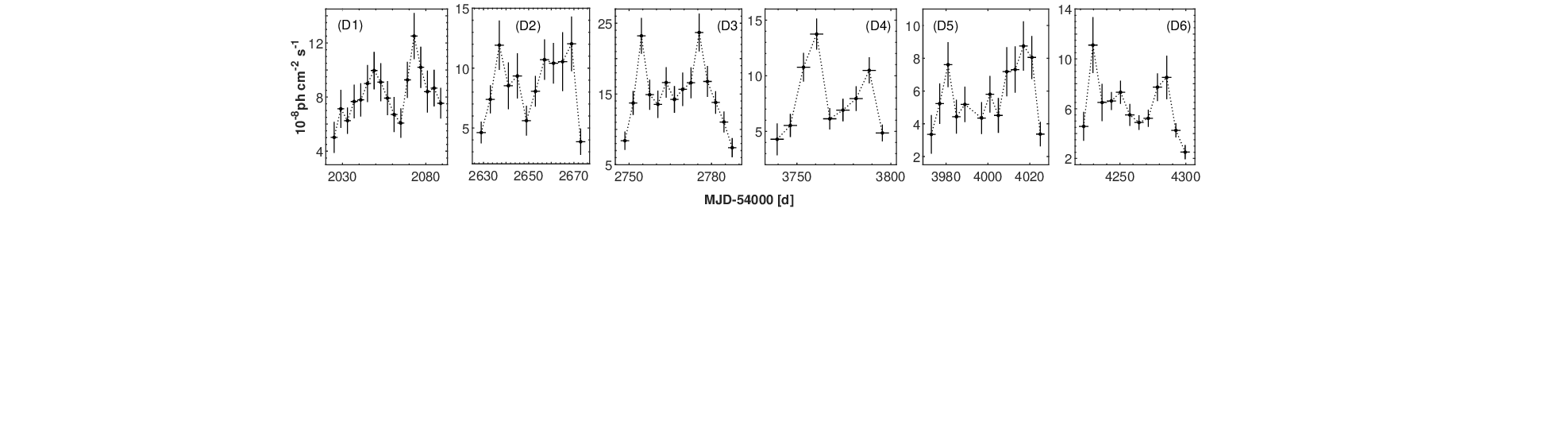}
\vspace{-0.8cm}  \caption{\label{sym} Examples of short-term LAT-band flares with different profiles: symmetric (A); positively (B) and negatively (C) asymmetric; two-peak (D).  The light curves of the different instances are constructed by means of the 0.3--300\,GeV flux values derived with 2-d to 1-week time integrations.}  \end{figure*}

Based on the maximum 0.3--300\,GeV flux and fractional variability amplitude from the weekly binned data, we have discerned six periods of the relatively stronger LAT-band activity compared to other time intervals from the entire 2008\,August--2023\,August period (see Table\,\ref{periods} for the summary). During these periods, the baseline flux level (defined by the curve connecting the lowest states of short-term flares) was generally higher than that in other epochs, showing a long-term increase and subsequent decline trends on a few months to yearly timescales. Below, we characterize each period in order of the descending  values of $F_{\rm max}$  and $F_{\rm var}$.  
 
\begin{table*}  \footnotesize  \begin{minipage}{170mm}
  \caption{\label{shortterm} Summary of shorter-term flares with the time ranges provided in Col.\,1 (extract). The maximum and mean 0.3--300\,GeV flux values (Columns 2 and 3, respectively) are given in units of 10$^{-8}$ph\,cm$^{-2}$s$^{-1}$; maximum-to-minimum flux ratio (Col.\,4); reduced Chi-squared and the corresponding degrees-of-freedom (Col.\,5) and fractional rms variability amplitude (Col.\,6; in percents); integration time used for the light curve construction  (Col.\,7). }   \centering    \begin{tabular}{cccccccccc}  \hline
 MJDs &  $F_{\rm max}$ & $F_{\rm mean}$ & $\Re$& $\chi^2_{\rm r}$/DOF  & $F_{\rm var}$ & Binning  \\
(1) & (2) & (3) & (4)& (5)&(6) & (7)  \\
    \hline
 \multicolumn{7}{c}{Symmetric flares}  \\
    \hline
(54)686--720&14.72(1.85)&	7.24(0.37)&	3.8(0.9)&3.11/13&28.7(5.4) &3\,d \\
(547)50--57&8.28(1.38)&4.91(0.53)	&2.1(0.5)&	6.92/2	&40.2(10.8) &4\,d \\
(548)46--52&10.57(1.54)&	5.81(0.76)&	6.9(2.1)&	6.75/2	&56.0(12.7) &3\,d\\
(550)22—47&7.72(1.20)&	4.83(0.34) &	3.0(0.8)&	4.76/8	&25.8(6.4) &3\,d \\
  \hline \end{tabular} \end{minipage}  \end{table*}

\subsection{Periods of strong LAT-band activity}

\begin{figure*}[ht!] 
\includegraphics[trim=6.3cm 0cm -1cm 0cm, clip=true, scale=0.89]{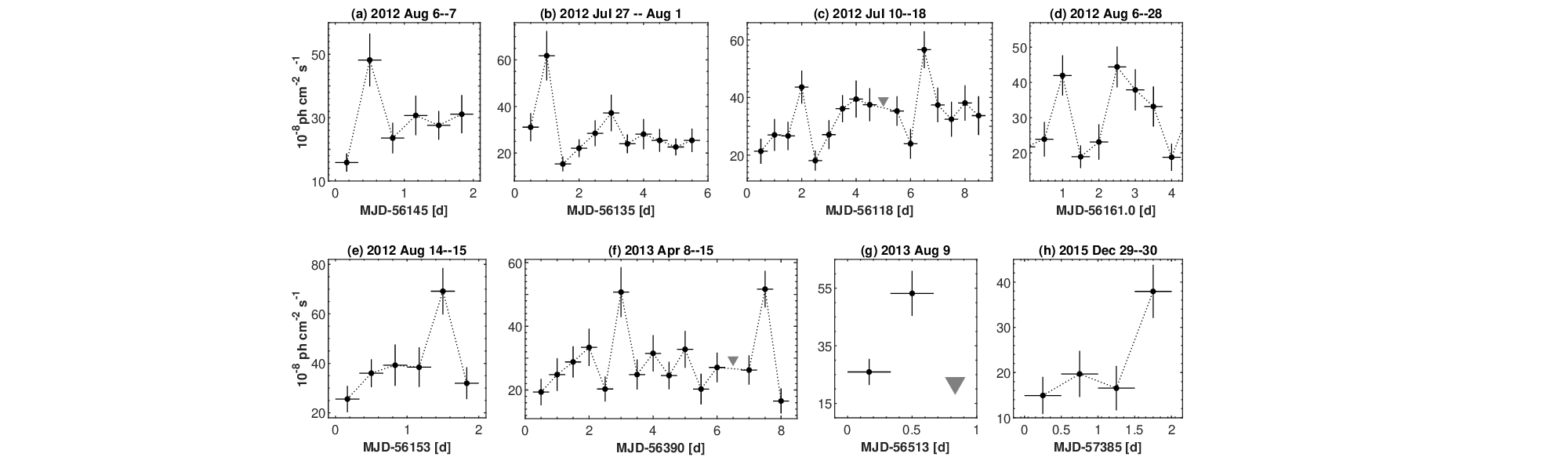} \vspace{-0.7cm}
 \caption{\label{idv} Selected 0.3--300\,GeV IDVs. The downward gray triangles correspond to the upper limits to the LAT-band flux when $TS$$<$9 and/or $N_{\rm pred}$$<$8.}  \end{figure*}

The source underwent the strongest LAT-band flaring activity in the second half of Period\,1, which lasted about 3 months (2012\,June--September, MJD\,(56)080--200; Figure\,\ref{per}A) and was characterized by the highest historical  0.3--300\,GeV flux in the case of the time integrations of 1\,d and longer (listed in Table\,\ref{frac}). This outburst was preceded by a plateau-like behaviour during $\sim$100\,d when the source showed minor fluctuations around the mean level of 9$\times$10$^{-8}$ph\,cm$^{-2}$s$^{-1}$. On the other hand, the latter was a factor of $\sim$2.5 higher than the  0.3--300\,GeV flux observed in the period's start and about 30\,per  cent higher than the mean level from all LAT observations of Mrk\,421 performed before 2012\,March. During the strongest LAT-band activity, the source showed a nearly-symmetric short-term flare by a factor of $\sim$2 lasting more than 2 weeks. In this period, the source  also underwent two short-term flares with a positive asymmetry (i.e., fast brightness increase and slower decline) and three other flares with a negative asymmetry (an opposite cadence; see Figure\,\ref{sym} for the corresponding examples and 
Table\,\ref{shortterm} for the maximum and mean flux values, maximum-to-minimum flux ratio and  $F_{\rm var}$). Finally, the source exhibited two other LAT-band flares of 1-2 months duration, showing a two-peak profile (see Figure\,\ref{sym}D for examples and Section\,5.2 for the corresponding physical implication). Moreover, 11 instances of the 0.3--300\,GeV flux doubling or halving occurred in the course of the aforementioned flares, and  the corresponding summary (maximum duration $\Delta t$, initial and final flux values, $F_{\rm var}$, doubling/halving timescale  $\tau_{\rm d,h}$).  Note that the quantity $\tau_{\rm d,h}$ was calculated as follows \citep{sa13}
\begin{equation}
 \tau_{\rm d,h}=\Delta t \times ln(2)/ln|(F_{\rm f}/F_{\rm i}),   
\end{equation}
with $F_{\rm i}$ and $F_{\rm f}$ to be the initial and final flux values, respectively, and $\Delta t$, the corresponding maximum duration. For these instances,  $\Delta t$=0.33--15\,d, yielding a range $\tau_{\rm d,h}$=0.21--9.65 days (see Table\,\ref{doubl} for details). The most extreme instance was recorded on  2012\,August\,6 (MJD\,56145) when the 0.3--300\,GeV brightness boosted by a factor of $\sim$3 in 8\,hr ($\tau_{\rm d}$=5.04\,hr; see Figure\,\ref{idv}a). Subsequently, the brightness declined by at least 29\% within the next 8\,hr (taking into account the flux error ranges; see also Table\,\ref{idvtable}). Another extreme instance with $\tau_{\rm h}$=12\,hr occurred during 2012\,July\,27--28 when the HE brightness dropped  by a factor of $\sim$4 within one day [MJD\,(56)135.5-136.5;  Figure\,\ref{idv}b)], and this behaviour was preceded by a flux increase by at least 38\% within the same time interval.

The source showed a brightness drop by a factor of $3$ within 16\,hr on August\,28--29 [MJD\,(561)67.67—68.33; see Table\,\ref{idvtable}]. A comparable drop occurred also during MJD\,(561)19.5--20.5 (2012\,July\,11-12; $\tau_{\rm h}$$\approx$14.5\,hr; Figure\,\ref{idv}c), which was preceded by a flux-doubling instance with $\tau_{\rm d}$=3.18\,d (see Table\,\ref{doubl}). The subsequent HE intraday variability (IDV) occurred on July\,16 (MJD\,56124) when the brightness increased by more than 70\% within the time interval shorter than 1\,d, followed by a flux halving during the next 3 days. A flux-halving instance with $\tau_{\rm h}$$\approx$14.5\,hr occurred also within MJD\,(5616)1.5--2.5 (2012\,August\,21--22), which was followed by a LAT-band IDV  after 1\,d  (see Figure\,\ref{idv}d and Table\,\ref{doubl}). Finally, another instance is detected from LAT data collected during the last 8-hr segment of 2012\,August\,15 (MJD\,56154) when the 0.3--300\,GeV brightness dropped by more than 35\% (Figure\,\ref{idv}e). However, the latter was also preceded by a flux-doubling instance with $\tau_{\rm d}$=1.71\,d.

\begin{table*}  \begin{minipage}{170mm}
\caption{\label{doubl} Summary of the 0.3—300 GeV flux doubling/halving instances (extract). Column (1) gives the maximum duration of the particular instance (in days). The initial and final flux values (in 10$^{-8}$ph\,cm$^{-2}$s$^{-1}$), reduced Chi-squared along with the corresponding degrees-of-freedom  and fractional rms variability amplitude (in percents) are  provided in Cols. 4--7, respectively.}
   \centering   \begin{tabular}{cccccccc} \hline
 $\Delta t$ (d) & MJDs &  Date(s) &$F_{\rm i}$ & $F_{\rm f}$ & $\chi^2_{\rm r}$/DOF  & $F_{\rm var}$   \\
  (1)	& (2) &	(3)&	(4) &	(5)&	(6)& (7)\\
    \hline
5&	(5470)3--7	&2008 Aug 25--29&	7.01(1.73)	&23.21(2.93)&	8.01/3	&45.5(8.7)\\
8	&(547)09--16	&2008 Aug 29—Sep 5	&15.94(2.02)&	5.39(1.10)	&11.37/3&	46.7(7.5)\\
8	&(548)29--36&	2008 Dec 29—2009 Jan 5&	2.67(0.75)&	9.16(1.36)	&6.63/3	&46.6(10.1)\\
6	&(550)17—22&	2009 Jul 5--10	&15.04(2.36)&	3.99(0.86)	&9.94/2	&62.8(12.8)\\
     \hline \end{tabular} \end{minipage}  \end{table*} 

Unfortunately, no XRT and UVOT observations were performed in the course of the LAT-band outburst of Mrk\,421, owing to the seasonal Sun restriction during June--October\footnote{See https://www.swift.ac.uk/sunpos.php}. The regular BAT-band observations yielded only two detections with 5$\sigma$ significance meanwhile, showing relatively low 15--150\,keV states of the source. However, an X-ray flare was detected with MAXI  in the epoch of the second 0.3--300\,GeV peak, in contrast to the first, higher peak (accompanied by enhanced radio-band activity; see the last panel of Figure\,\ref{per}A). Moreover, that flare was relatively moderate compared to those detected in Periods  2  and 4--5 (see Figures \ref{per}B--\ref{per}C and \ref{per}E). As for the time interval with the plateau-like HE behaviour of the source, the XRT observations mostly showed low 0.3--10\,keV states (see the 3-rd panel in Figure\,\ref{per}A). A strong flare was observed in the \emph{UVW1--UVW2} bands around MJD\,56075, which was weaker in the optical \emph{V}-band [panels (e)--(g)]. Finally, the source was not observed with FACT or other Cherenkov-type telescopes during the entire period. 

The source showed another LAT-band outburst in 2013\,March--April [MJD\,(56)350--410, Period\,2], when the exceptionally strong 0.3--10\,keV flare occurred (Figure\,\ref{per}B). A similar behaviour was observed also with FACT and UVOT.  However, Mrk\,421 was relatively quiet in the hard X-ray bands (as observed by BAT and MAXI), and a stronger MAXI-band activity was recorded about 1\,month later [panel (e)]. Note that the relatively high 0.3--300\,GeV flux (exceeding the threshold of 10$^{-7}$ph\,cm$^{-2}$s$^{-1}$) was observed during most parts of Period\,2. On weekly timescales, the source showed five symmetric flares with $F_{\rm var}$=25--29\,per cent. Moreover, this period was characterized by five positively and one negatively-asymmetric instances, as well by three two-peak flares (see Table\,\ref{shortterm}). Note that the short-term 0.3--300\,GeV flare in the epoch of the X-ray outburst showed a positive asymmetry, whereas it was preceded by the symmetric and positively-asymmetric flares, respectively. The source underwent another strong UV-flare centered on MJD\,56287 and  showing no corresponding LAT-band, X-ray and optical "counterparts". Nevertheless, the strongest optical \emph{V}-band flare was observed at the period's end, when the 0.3--300\,GeV brightness was showing a long-term declining trend.

The source also showed 16 instances of the LAT-band flux doubling/halving with  $\tau_{\rm d,h}$=0.6--7.4\,d and  $F_{\rm var}$=32.6(5.4)--74.0(10.6)\,per cent during that period (Table\,\ref{doubl}). The most extreme variability was observed on 2013\,April\,15 (MJD\,56397), when  the 0.3--300\,GeV flux dropped by a factor of $\sim$3 within the 14.5\,hr time interval after the  interval's highest brightness (Figure\,\ref{idv}f). This instance was preceded by another IDV, incorporating a brightness increase by more than 50\% (see  Table\,\ref{idvtable}). Three other IDVs were  respectively characterized by: (1) a brightness drop by at least 32\% within the first half of 2013\,March\,26 (MJD\,56377; Table\,\ref{idvtable}); (2) a brightening by $\sim$80\% and subsequent drop to the initial level, with the entire cycle lasting 1.5\,d [April\,10-11; MJD\,(5639)2.0--3.5; Figure\,\ref{idv}f]; (3) a brightness increase by more than 64\%  within the first half of 2013\,August\,9 (MJD\,56513). Note that this period included the aforementioned two 1-hr robust detections of the source with the 0.3--300\,GeV flux of $\approx$10$^{-6}$ph\,cm$^{-2}$s$^{-1}$. 

Very strong 0.3--300\,GeV activity was recorded also in the middle  of Period\,5, reaching a level of 3$\times$10$^{-7}$ph\,cm$^{-2}$s$^{-1}$ during the negatively-asymmetric flare peaking on MJD\,58166 (see Figures \ref{per}C and \ref{sym}Cf). This instance was preceded by another short-term flare with a negative asymmetry, whereas two flares with opposite asymmetry, two symmetric  and three two-peak flares occurred in the same period (see  Table\,\ref{shortterm}). At the onset of the aforementioned highest-peak flare, the source also underwent a very strong VHE outburst by a factor of $\sim$10, which showed a peak brightness about one month earlier than the LAT-band flare. Note that the peak VHE brightness during this instance was comparable to that observed in 2013\,April. There was another very strong FACT-band  flare by a factor of 8, peaking on MJD\,58111 (in 8\,d) and followed by a very fast drop to the initial brightness (in 2\,d). During this time window, only a low-amplitude HE flare was recorded with LAT. However, the source showed a strong $\gamma$-ray flare in the both LAT and FACT bands around MJD\,58200, when the VHE emission boosted by a factor of $\sim$3. This flare was not accompanied by the comparable XRT-band activity, in contrast to the previous VHE instances when the  0.3--10\,keV flux exceeded the threshold of 150\,cts\,s$^{-1}$ (the second highest level after the exceptionally strong X-ray outburst in 2013\,April). These fast instances were superimposed on the long-term flare lasting $\sim$4\,months, which was also recorded  with LAT, BAT, MAXI and UVOT. However, the source showed lower optical brightness in the epoch of the highest UV to VHE states, and it underwent a  flare during the long-term  declining trend in the LAT band. Note also the strongest \emph{V}-band flare and elevated 43\,GHz in the period's start when the source underwent a negatively-asymmetric LAT-band flare by a factor of $\sim$4. In this period, the source showed a flux doubling/halving ten times, generally occurring during the aforementioned short-term LAT-band flares (Table\,\ref{doubl}). During one  of these instances, the source underwent a 0.3--300\,GeV IDV incorporating a flux increase by more than 78\%  within the first half of 2018\,February\,9 (MJD\,58158; see Table\,\ref{idvtable}).

In Period\,3, the source was detected mainly above the 0.3--300\,GeV brightness level of 10$^{-7}$ph\,cm$^{-2}$s$^{-1}$, by showing one symmetric and three two-peak flares  (Figure\,\ref{per}D). Moreover, one observed four and three short-term flares with positive and negative asymmetries, respectively (Table\,\ref{shortterm}). Meanwhile, two fast, strong VHE flares were detected with FACT (around MJD 56228 and 562741, respectively). The second instance occurred at the onset of the strongest LAT-band flare, while Mrk\,421 was exhibiting only a minor 0.3--10\,keV one.  On the other hand, the source was detected only 1-2 times with 3$\sigma$ significance by FACT during the two strongest XRT flares (centered on MJD 56639 and 56702, respectively; the consecutive, negatively-asymmetric LAT-band flares were presented meanwhile). Strong UV activity was observed during the short-term LAT and XRT band flares. However, the source did not exhibit strong X-ray flares in the MAXI and BAT bands. Similar to the above discussed periods, (i) the optical and radio variabilities were even less well-correlated with the LAT-band flares; (ii) the detected eleven flux doubling/halving instances were associated with the aforementioned LAT band flares. However, no 0.3--300\,GeV IDVs occurred in that period.

\begin{table*}[ht!]  \begin{minipage}{170mm}
  \caption{\label{idvtable} Summary of the 0.3--300 GeV IDVs. The initial and final 0.3--300\,GeV flux values (Cols 3 and 4, respectively) are given in units of 10$^{-8}$ph\,cm$^{-2}$s$^{-1}$; Col.\,(4) presents the reduced Chi-squered and the corresponding degrees-of-freedom; Col.\,5: fractional rms variability amplitude (in percents).}       \centering  
  \begin{tabular}{ccccccc}  \hline
Date/MJD   &$F_{\rm i}$ & $F_{\rm f}$ & $\chi^2_{\rm r}$/DOF  & $F_{\rm var}$ \\
 (1)	& (2) &	(3)&	(4) &	(5)\\     \hline
2012 Jul 11/(56119.)25--50	&	18.81(3.73)&44.37(6.72)&11.06/1&	54.6(12.4)\\
2012 Jul 11--12/(561)19.5-20.5	&44.22(5.85)&14.11(3.05)&15.53/1&	74.1(10.8)\\
2012 Jul 15--16/(561)23.5-24.5	&	23.92(5.17)&56.60(6.72)&14.86/1&	55.4(10.7)\\
2012 Jul 16/56124&23.92(5.15)&56.60(6.50)&	11.52/1&	47.9(10.7)\\
2012 Jul 27/(56135).0--5&	31.11(5.69)&	61.79(11.8)&10.89/1&	84.7(16.3)\\
2012 Jul 27--28/(561)35.5-36.5&61.79(11.8)&	14.49(3.16)&14.99/1&	84.7(16.3)\\
2012 Aug 6/(56145.)33--67&	15.83(2.80)&	48.20(8.29)&13.66/1&	68.8(13.9)\\
2012 Aug 6/(56)145.67--166.00&48.20(8.29)&23.83(4.66)&10.95/1&	36.7(10.5)\\
2012 Aug 15/(561)54.67-55.0&68.66(9.5)&	26.26(5.55)&14.85/1&	61.0(11.8)\\
2012 Aug 22--23/(561)61.5-62.5	&42.45(6.07)&14.68(3.20)&16.38/1&	66.6(12.2)\\
2012 Aug 23--24/(561)62.5-63.5&	22.10(5.01)&45.92(5.84)&10.85/1&	49.5(11.5)\\
2012 Aug 28--29/(561)67.67—68.33&48.30(9.09)&16.21(3.26)&11.03/1&	67.1(15.3)\\
2013 Mar 26/(56377.)00--50&	58.95(9.72)&	23.63(9.80)&	10.99/1&	59.1(1.35)\\
2013 Apr 10/56392&	20.28(3.91)&	50.73(7.76)&	13.46/1&	58.1(1.30)\\
2013 Apr 10-11/(56)392.5-393.5&50.73(8.76)&	20.30(4.19)&10.93/1&	45.2(11.8)\\
2013 Apr 14--15/(56)396.5-397.5&26.26(4.75)&51.66(6.10)&10.96/1&	43.9(10.2)\\
2013 Apr 15/(56397.)417--500&106.38(18.54)&	36.61(7.79)&	12.04/1&	66.1(14.4)\\
2013 Aug 9/(56513.)00--50&	24.92(4.30)&	54.20(7.79)&	10.85/1&	49.9(11.5)\\
2015 Dec 30/57386	&15.05(3.51)&	39.93(6.65)&	10.91/1&	61.0(14.0)\\
2016 May 18/57526	&12.03(2.82)&	31.34(4.8)&	12.03/1&	60.3(13.1)\\
2018 Feb 9/(58158.)0--5	&26.19(4.82)&	66.49(11.12)&	11.04/1&	57.9(13.4)\\
2022 May 8/59707.5--59708.0&24.12(5.54)&	4.85(0.96)&	11.75/1	&	90.0(19.8)\\
2022 May 15/59714.5--59715.0&15.65(3.75)&55.74(11.43)&	10.99/1&	75.7(17.4)\\
2022 Oct 9/59861&	24.45(3.71)&55.74(11.43)&	10.89/1	&	54.9(12.6)\\
2022 Nov 30/59913.33--59914.00&	14.08(4.14)&	38.31(6.03)&	10.97/1	&	62.4(14.3)\\
   \hline \end{tabular} \end{minipage} \end{table*}

A strong MEV--GeV band flux variability was recorded also in the subsequent period, including two symmetric, four two-peak, eight positively and three negatively asymmetric flares  (Figure\,\ref{per}E). These instances were superimposed on the gradually declining baseline 0.3--300\,GeV brightness level, and this trend was observed during almost 2\,years. Note also that the source showed  frequent detections with FACT, incorporating seven strong VHE flares. The first, strongest instance coincided with those recorded with LAT, BAT and MAXI (no contemporaneous Swift observations were carried out). Note that the source was not detected with FACT in the epoch of the highest XRT and MAXI-band peak. On the contrary, the strong VHE flare around MJD\,57506 was not accompanied by the corresponding 0.3--10\,keV activity. However, nearly simultaneous strong flares were observed with LAT, FACT, XRT and BAT. Note also that UV-band peaks had no comparable "counterparts" in the higher-energy bands, except for the flare occurring around MJD\,57390. In the latter case, the strongest long-term \emph{V}-band flare also showed a peak. The source showed 23 instances of the 0.3--300\,GeV flux doubling/halving with $\tau_{\rm d,h}$=2.37--16.01\,d and $F_{\rm var}$=24.6(5.8)--73.2(9.5)\,per cent. The largest-amplitude instance [brightness drop by a factor of $\sim$3 during (573)87—90] was preceded by the 0.3--300\,GeV IDV incorporating a brightness increase by $\sim$80\% within 1\,d (Tables\,\ref{doubl} and Figure\,\ref{idv}h). The second IDV from this period was characterized by a similar amplitude but showed an opposite cadence (see Table\,\ref{idvtable}). 

Finally, Period\,6 was also notable for the target's enhanced MeV-GeV activity: a general elevated base state, lasting  almost 14\,months,  was superimposed  by three symmetric, four two-peak, three positively and eight negatively asymmetric flares  (Figure\,\ref{per}F).  Five strong X-ray flares were recorded with the XRT, and each one was accompanied by the 0.3--300\,GeV "counterparts". However, X-ray flaring activity was relatively moderate in the MAXI-band, and only four detection at the 5$\sigma$ confidence level was recorded with BAT. The first, second and third XRT-band flares were accompanied by those in the optical--UV bands, followed by a long-term declining trend until the period's end, and no remarkable activity was observed along with the last two X-ray flares. The first short-term LAT-band flare was accompanied by the highest 43\,GHz  state of the source. Among 18 instances of the LAT-band flux doubling/halving with $\tau_{\rm d,h}$=1.06--12.96\,d, the first one incorporated a 0.3--300\,GeV IDV with a factor of $>$2 boost in the MeV--GeV brightness within the second half of MJD\,59714 (see  Tables\,\ref{doubl} and \ref{idvtable}). Another extreme IDV (brightening by $\sim$80\% within 16\,hr on MJD\,59913) was followed by a flux halving instance with $\tau_{\rm h}$=1.50\,d. The third IDV (brightening by more than 55\% on  MJD\,59861) was a part of the short-term flare with a negative asymmetry (see Table\,\ref{shortterm}). Note that this period was characterized by the highest-amplitude flux doubling/halving instance during 2008--2023 [$F_{\rm var}=$95.8$\pm$18.7\,per cent, factor of $\sim$5 drop during MJD\,(5994)3--6].

\begin{table*}  \begin{minipage}{170mm} \tabcolsep 3.3pt \centering
  \caption{\label{distrtable} Distribution of the photon indices from the  power-law  and  logparabolic 0.3--300\,GeV spectra, corresponding to the different time integrations (from two weeks down to 1\,d) during 2008--2023  and periods (bottom, only for the $\Gamma$-index). The maximum, minimum, mean and distribution peak values are presented in the first, second, 3rd and 4th rows for the both parts of the table, respectively.} \vspace{-0.4cm}  
  \begin{tabular}{cccccccccccccccc}    \\  \hline  
     &   \multicolumn{9}{c}{2008--2023}\\ 
  \hline
  & $\Gamma$(2\,w) &$\Gamma$(1\,w) & $\Gamma$(4\,d)& $\Gamma$(3\,d)  & $\Gamma$(2\,d) &$\Gamma$(1\,d)& $\alpha$(2\,w)  & $\alpha$(1\,w) &  $\alpha$(4\,d)  \\
    \hline
 Max. &2.55(0.18) &2.59(0.26) & 2.83(0.38)&2.89(0.28) & 2.90(0.30)&2.86(0.28) &1.98(0.07)  &2.44(0.20) &2.62(0.20)   \\
Min. &1.56(0.09) & 1.37(0.15)&1.31(0.09) &1.24(0.13) &1.24(0.12) &1.21(0.13) & 1.18(0.07) &1.13(0.10) &1.05(0.13)  \\
Mean &1.80(0.01) &1.80(0.01 & 1.80(0.01)&1.79(0.01) &1.82(0.01) &1.82(0.01) &1.57(0.02) &1.63(0.01) &1.58(0.01)  \\
Peak &1.79(0.01) &1.76(0.01) &1.77(0.01) &1.76(0.01)&1.76(0.01) & 1.75(0.01)&1.55(0.02) &1.52(0.01) & 1.50(0.01) \\   \hline
&\multicolumn{3}{c}{2008--2023}&\multicolumn{6}{c}{$\Gamma$}\\ \hline
& $\alpha$(3\,d) & $\alpha$(2\,d)& $\alpha$(1\,d)& Per\,1& Per\,2  & Per\,3 &  Per\,4 &  Per\,5& Per\,6 \\      \hline  
Max. &2.39(0.20)& 2.61(0.25)&2.60(0.23)&2.31(0.19) &2.38(0.22) &2.48(0.25) &2.71(0.25) &2.64(0.29)& 2.34(0.23) \\
Min.&1.05(0.13) &1.10(0.14) &1.03(0.17) &1.51(0.07) &1.41(0.07) &1.30(0.11) &1.24(0.11) &1.30(0.14)& 1.29(0.13)\\
Mean&1.61(0.01) & 1.56(0.01)&1.73(0.01 &1.82(0.01) & 1.75(0.01) &1.82(0.01) &1.76(0.01) &1.84(0.01)&1.83(0.01) \\
Peak&1.50(0.01) &1.51(0.01) &1.57(0.01) &1.76(0.01) &1.74(0.01) &1.77(0.01) &1.78(0.01) & 1.82(0.01)&1.78(0.01) \\
 \hline     \end{tabular} \end{minipage} \end{table*}

 \section{Spectral Results}

\begin{figure*}[ht!] 
\includegraphics[trim=6.0cm -0.3cm -0.9cm 0cm, clip=true, scale=0.89]{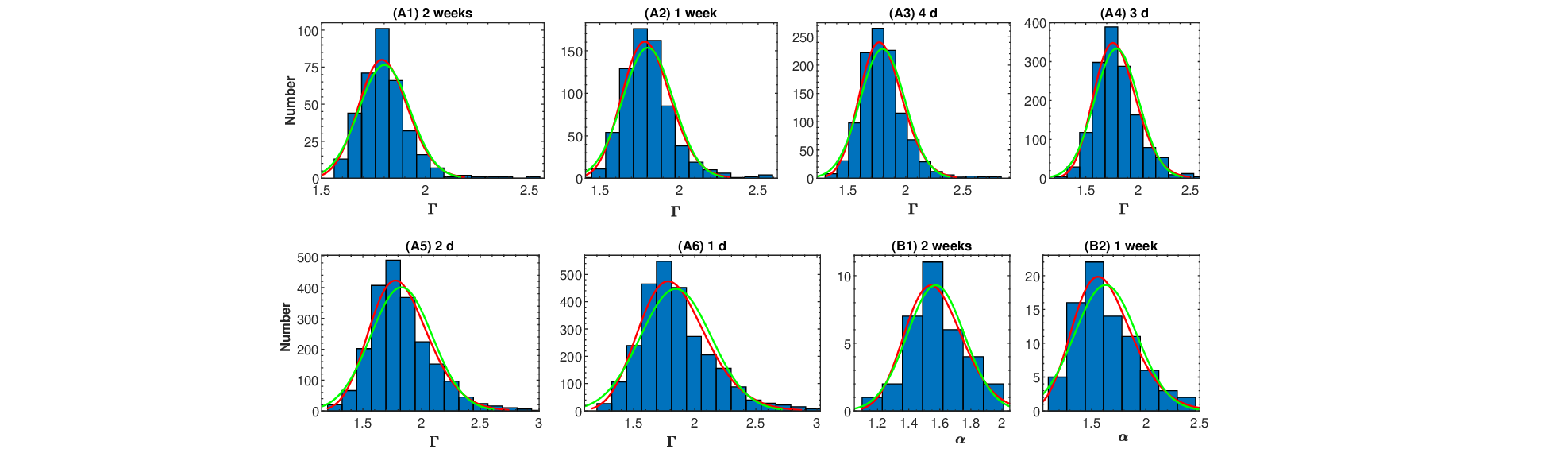}
\includegraphics[trim=6.0cm 4.9cm -0.9cm 0cm, clip=true, scale=0.89]{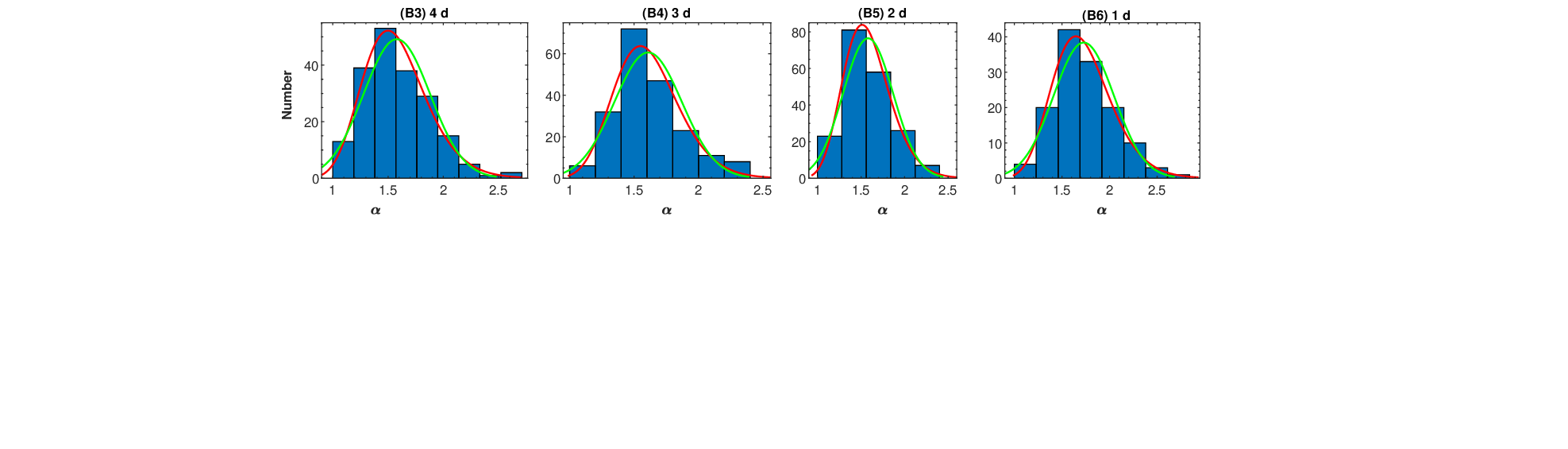} \vspace{-0.7cm}
 \caption{\label{inddistr} Distribution of the photon indices from the power-law and logparabolic spectra during 2008--2023, derived from the 0.3--300\,GeV data by using different time integrations. The red and green curves show the lognormal and Gaussian fits to the histograms, respectively.}  \end{figure*}

The distribution of the $\Gamma$-values obtained from the power-law spectral analysis for the entire 2008--2023 period is presented in Figures\,\ref{inddistr}A1--\ref{inddistr}A6, each histogram representing the results obtained from the different time integration from two weeks down to one day. The distribution details (maximum, minimum, mean and peak values)  are provided in Table\,\ref{distrtable}. The latter demonstrates that the source showed a very large spectral variability from the extremely hard $\Gamma$-values ($\Gamma$$\lesssim$1.5) to extremely soft ($\Gamma$$\gtrsim$2.5) spectra. Namely, the overall range of the photon index $\Delta \Gamma$=0.99(0.20) for the two-weekly binned LAT data and $\Delta \Gamma$=1.65(0.31) in the case of the 1--3 day integrations. The distribution peaks are derived by fitting the corresponding histogram with the lognormal function, which generally showed slightly better statistics compared to other (e.g., Gaussian) functions (see Table\,\ref{logntable}). 

\begin{figure*}[ht!] 
\includegraphics[trim=6.0cm 4.8cm 0cm 0cm, clip=true, scale=0.88]{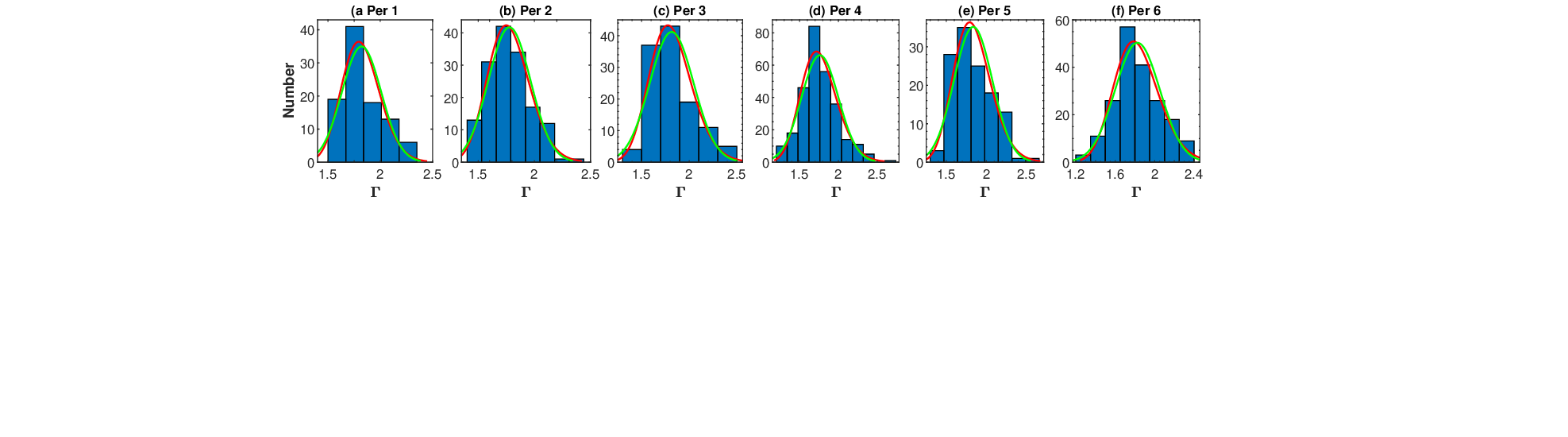} 
\includegraphics[trim=6.3cm 4.8cm 0cm 0cm, clip=true, scale=0.9]{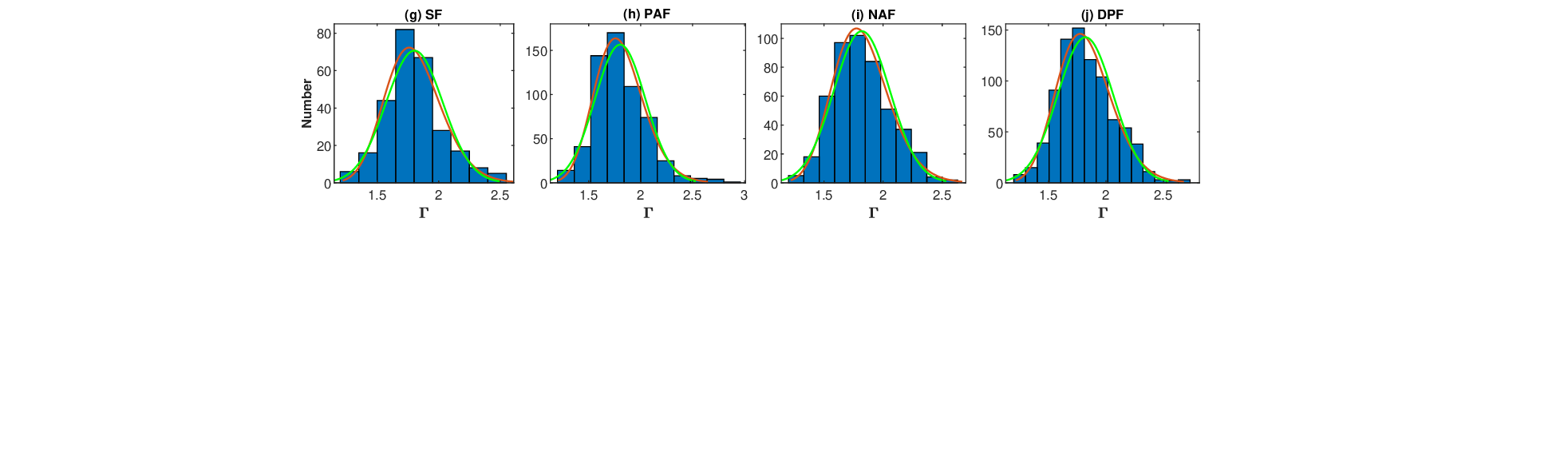} \vspace{-1cm} 
\caption{\label{inddistr1} Distribution of the photon-index  $\Gamma$ in different periods [panels (a)--(f)],  as well as for those time intervals showing the LAT-band flares of Mrk\,421 with different profiles [panels (g)--(j); SF -- symmetric flares, PAF -- positively-asymmetric flares, NAF  -- negatively-asymmetric flares and DPF -- double-peak flares].
The red and green curves show the lognormal and Gaussian fits to the histograms, respectively. }
 \end{figure*}

The peak and mean values from the histograms, corresponding to the different time integrations during the entire 2008--2023 period, are close to each other ($\Gamma_{\rm p}$=1.75(0.01)--1.79(0.01), $\Gamma_{\rm mean}$=1.79(0.01)--1.82(0.01). Note that the $\Gamma$-values from the time bins corresponding to the non-robust detections of the source are not included in the distribution study. Since the percentage of such bins are gradually higher with shorter integration times, the distribution from the latter are biased towards the higher MeV--GeV states of  Mrk\,421, and the slight differences between the $\Gamma_{\rm p}$ and $\overline{\Gamma}$ values could be primarily due to this selection effect.  The distributions of the photon index from the aforementioned six periods are relatively more different (see the bottom part of Table\,\ref{distrtable} and Figures\,\ref{inddistr1}a--\ref{inddistr1}f): the ranges of the peak and mean values are  $\Gamma_{\rm p}$=1.73(0.01)--1.82(0.01) and  $\Gamma_{\rm mean}$=1.75(0.01)--1.84(0.01), respectively. On average, the 0.3--300\,GeV spectra showed the hardest distribution peak in Period 1 which was characterized by  one the strongest flaring activity of the source (see  Section\,3.1).

As noted above, the LAT-band power-law spectra of Mrk\,421 were sometimes extremely hard with $\Gamma$$\lesssim$1.5, and this happened within the different time intervals ranging from intraday to 25 days (see Table\,\ref{hard}). Note that each period contained a different number of  extremely hard 0.3--300\,GeV spectra (see Table\,\ref{hard} and Section\,5.3 for the corresponding physical implications). Moreover, we checked the 4-weekly binned LAT data for  possible hardening beyond 10\,GeV (see Section\,5.3 for the discussion in the context of hadronic cascades). This integration time was required to achieve the target's robust detection in the 10--300\,GeV energy range for the most of the time bins. Table\,\ref{10gev} presents 23 cases of spectral  hardening beyond 10\,GeV with $\Gamma$=0.93$\pm$0.20 to $\Gamma$=1.65$\pm$0.16.

\begin{table}[ht!] \centering \footnotesize  \tabcolsep 1.5pt   \begin{minipage}{80mm}
  \caption{\label{logntable}  Goodness-of-fit for the flux and photon-index distributions in various energy bands. The distribution fits are performed by using the Gaussian and lognormal functions  (Columns 2 and 3, respectively). The acronyms SF, PAF, NAF and DPF have the same meanings as in  Figure\,\ref{inddistr1}.}
  \begin{tabular}{ccccccccc}     \hline \centering
Energy Band & $\chi^2_{\rm r}$/DOF(Gauss)  & $\chi^2_{\rm r}$/DOF(Logn) \\
(1) & (2) & (3) \\   \hline
 $\Gamma$ (2\,w) & 1.13/14&0.99/14   \\
 $\alpha$ (2\,w) & 1.06/6&0.94/6   \\
 $\Gamma$ (1\,w) & 1.14/14&1.03/12   \\
 $\alpha$ (1\,w) & 1.29/8&1.07/8   \\
 $\Gamma$ (4\,d) & 1.16/14&1.06/14   \\
  $\alpha$ (4\,d) & 1.27/8&1.02/8  \\
 $\Gamma$ (3\,d) & 1.15/11&1.05/11  \\
 $\alpha$ (3\,d) & 1.22/6&1.08/6  \\ 
 $\Gamma$ (2\,d) & 1.19/13&1.08/13   \\
  $\alpha$ (2\,d) & 1.27/6&1.16/6   \\
 $\Gamma$ (1\,d) & 1.29/15&1.10/15   \\
 $\alpha$ (1\,d) & 1.24/7&1.06/7   \\
  $\Gamma$ (2\,d,\,Per\,1) & 1.21/4&1.10/4   \\ 
  $\Gamma$ (2\,d,\,Per\,2) & 1.20/7&1.09/7   \\ 
  $\Gamma$ (2\,d,\,Per\,3) & 1.12/5&0.96/5   \\ 
 $\Gamma$ (2\,d,\,Per\,4) & 1.22/10&0.97/10   \\  
   $\Gamma$ (2\,d,\,Per\,5) & 1.22/7&1.15/7   \\  
 $\Gamma$ (2\,d,\,Per\,6) & 1.20/7&1.10/7   \\  
   $\Gamma$ (2\,d,\,SF) & 1.21/10&1.12/10   \\ 
 $\Gamma$ (2\,d,\,PAF) & 1.18/13&1.04/13   \\    
  $\Gamma$ (2\,d,\,NAF) & 1.22/12&1.14/12   \\ 
  $\Gamma$ (2\,d,\,DPF) & 1.24/16&1.07/16   \\  
 $F_{\rm 0.3--300\,GeV}$ (2\,w)  & 1.39/10&0.98/10   \\
 $F_{\rm 0.3--300\,GeV}$ (1\,w) & 1.43/10&1.02/10   \\
$F_{\rm 0.3--300\,GeV}$ (4\,d) & 1.55/13 &1.07/13   \\
$F_{\rm 0.3--300\,GeV}$ (3\,d) & 1.55/14&1.09/14   \\
$F_{\rm 0.3--300\,GeV}$ (2\,d) & 1.88/14&1.06/14   \\
$F_{\rm 0.3--300\,GeV}$ (1\,d) & 1.71/13&0.95/13   \\
$F_{\rm 0.3--300\,GeV}$ (2\,d,\,Per\,1) & 2.36/9&1.11/9 \\
$F_{\rm 0.3--300\,GeV}$ (2\,d,\,Per\,2) & 1.45/6&0.96/6 \\
$F_{\rm 0.3--300\,GeV}$ (2\,d,\,Per\,3) & 1.39/10&0.98/10 \\
$F_{\rm 0.3--300\,GeV}$ (2\,d,\,Per\,4) & 1.36/9&1.13/9 \\
$F_{\rm 0.3--300\,GeV}$ (2\,d,\,Per\,5) & 1.35/9&1.12/9 \\
$F_{\rm 0.3--300\,GeV}$ (2\,d,\,Per\,6) & 1.66/7&1.16/7 \\
$F_{\rm 0.3--300\,GeV}$ (2\,d,\,SF) & 1.48/10&0.97/10 \\
$F_{\rm 0.3--300\,GeV}$ (2\,d,\,PAF) & 1.54/11&1.10/11 \\
$F_{\rm 0.3--300\,GeV}$ (2\,d,\,NAF) & 1.37/10&1.15/11 \\
$F_{\rm 0.3--300\,GeV}$ (2\,d,\,DPF) & 1.33/10&1.01/16 \\
\hline \end{tabular} \end{minipage} \end{table}

The parameter $\Gamma$ showed a strong variability on various timescales, as shown in Figure\,\ref{hist} where the timing behaviour of the  photon index is presented by different time integrations (similar to the 0.3--300\,GeV photon flux). First of all, these instances were related to the emergence of the extremely hard spectra. The most extreme spectral hardening/softening instances (the largest and/or fastest) are presented in Table\,\ref{hard}. The largest-amplitude hardening by $\Delta \Gamma$=1.34(0.12) occurred during MJD\,(562)24--26 (Period\,5, in the epoch of one of the strongest LAT-band flaring activity), and the instance with the comparable amplitude (although occurring within 16\,d) was observed after the largest softening by $\Delta \Gamma$=1.60(0.33) within 18\,d [MJD\,(59)385--402].  Note that hardenings/softenings by $\Delta \Gamma$$>$1 (taking into account the error ranges) during 2-5 days was observed more than 10 times. For example, the spectrum showed a subsequent softening and hardening by $\Delta \Gamma$=1.30(0.28) and $\Delta \Gamma$=1.12(0.29), respectively, within the 5-d time interval [(561)08--12, in Period\,5]. 

As noted above, only a minority of the 0.3—300 GeV spectra (5\%  to 13\% with the different time integrations) showed a spectral curvature  with the significance 2$\sigma$ and higher. Namely, the lowest percentage is associated with the 2-day binned data, while the highest one  -- with the 4\,d integrations. Note that the reference energy $E_0$ was generally close to 1.286\,GeV when leaving this parameter free during the spectral fit fit the logparabolic model and, consequently, we re-fitted the curved spectra by fixing $E_0$ to this value (in order to minimize the uncertainties related to the photon index $\alpha$ and curvature parameter $\beta$). The distribution of the $\alpha$-values from the different time integrations during the entire 2008--2023 period is presented in Figures \ref{inddistr}B1--\ref{inddistr}B6, and the corresponding distribution properties are provided in Table\,\ref{distrtable}. As far as we see, the photon-index range is  relatively narrow ($\Delta \alpha$=0.80(0.10) in the case of 2-weekly binned data, and it increases to $\Delta \alpha$=1.57(0.24) in the case of the 4-d integration. Similar to the $\Gamma$-index, the corresponding histograms are relatively well-fitted with the lognormal model (except for the two-week integration; see also Table\,\ref{distrtable}), and the $\alpha$-values were sometimes extremely hard down to $\alpha$=1.03$\pm$0.13. Note that the logparabolic LAT-band spectra showed a concentration within some time intervals. Finally, the curvature parameter showed a range between $\beta$=0.12$\pm$0.06 (two-weekly binned data from the flaring epochs) to $\beta$=1.64$\pm$0.22 (1-d time integration).

\section{DISCUSSION AND CONCLUSIONS}
\subsection{Variability Character}
Intense studies of the MWL flux variability pattern provide us with one of the most efficient tools for drawing conclusions about the unstable processes occurring  in blazars. Especially important objective is related to the detection of the periodic flux  variability, which can be associated with the jet precession (see, e.g., \citealt{tav08,sob17}). Namely,  there should be a primary BH with a jet pointed to the observer and accretion disc, while the secondary, smaller-mass BH is moving on a highly eccentric orbit around the system's mass center. The dominant effect is simply an imprint of the primary BH's orbital velocity on the jet, causing the jet's viewing angle to vary by the greatest amount, and the highly-relativistic ejected material is expected to have the same velocity component in the observer’s rest frame. Consequently, the jet will precess with respect to the distant observer, and one should observe a periodic flux variability. 

Generally, the clear identification of periodic variability of blazar is problematic owing to the complexity of  light curves and the lack of data trains large enough to provide an adequate sampling over large time intervals. The long-term regular LAT observations are especially valuable in this regard. In order to detect a possible periodical behavior of Mrk\,421 from these observations, we constructed the Lomb-Scragle (LSP) periodogram \citep{l76,s82}, representing an improved Fourier-based technique which is designed for unevenly-sampled time series $g_n$ without interpolation for the data gaps \citep{v18}: \vspace{-0.2cm}
\begin{equation} \begin{array}{c l}
 P(f)={A^2\over 2}\left(\sum\limits_{n}{g_n cos(2\pi f[t_n-\tau]} \right)^2+\\
 {B^2\over 2}\left(\sum\limits_{n}{g_n sin(2\pi f[t_n-\tau]} \right)^2,
 \end{array}  \vspace{-0.2cm} \end{equation}
where \emph{A}, \emph{B}, and $\tau$ are the arbitrary functions of the frequency \emph{f} and observing times $\{t_i\}$.  The LSP exhibits the most significant spectral power peak in the case of the periodicity existence, and estimates its significance by testing the false alarm probability of the null hypothesis.

\begin{table*}  \footnotesize  \begin{minipage}{170mm}
  \caption{\label{hard} Extremely hard  power-law  spectra (extract). For each photon index value, the corresponding time range in MJD and UTC,  $N_{\rm pred}$ and TS values are presented.}    \centering 
  \begin{tabular}{ccccccccccc}  \hline
 Dates/MJDs& $N_{\rm pred}$& TS & $\Gamma$ & Dates/MJDs&  $N_{\rm pred}$& TS & $\Gamma$ \\    \hline
 2008 Aug 19--20/(546)98--99	&10	&81	&1.39(0.12)	&2010 Oct 30—31/(55)499—500&	14	&115&	1.48(0.10)\\  
 2008 Aug 25--29/(5470)4--5&	44&	235&	1.37(0.08)&	2010 Nov 29—30/55529&	9&	62&	1.28(0.13)\\
2008 Sep 22--23/(5473)1—2	&11&	95&	1.24(0.11)&	2011 Jan 8--9/(555)69—70&	19&	162	&1.31(0.09)\\
2008 Sep 28--29/(5473)7—8&	9	&80	&1.28(0.11)&	2011 Feb 8--15/(5560)0--7&	56&	310	&1.53(0.05)\\
     \hline \end{tabular} \end{minipage} \end{table*}  
     
Since every periodicity searching technique requires at least one alternative check, we adopted the weighted wavelet Z-transform (WWZ) method (\citealt{f96} and references therein), which performs a periodicity analysis in both the time and frequency domains. It is defined as follows \vspace{-0.1cm}
\begin{equation}
WWZ={(N_{\rm eff}-3)V_y\over{2(V_{\rm x}-V_{\rm y})}},
\vspace{-0.1cm} \end{equation}
with $N_{\rm eff}$, the so-called effective number of data points; $V_{\rm x}$ and $V_{\rm y}$, the weighted variation of the data $x(t)$ and model function $y(t)$, respectively. The WWZ is based on the
Morlet wavelet \citep{g84} $ f(z)=e^{-cz^2}(e^{iz}-e^{-1/4c})$, where the constant $e^{-1/4c}$ is selected in the manner the wavelet's mean value to be zero. Similar to the LSP, the WWZ technique is robust against missing data. 

\begin{figure*}[ht!] 
\includegraphics[trim=6.2cm 2.8cm 0.2cm 0cm, clip=true, scale=0.91]{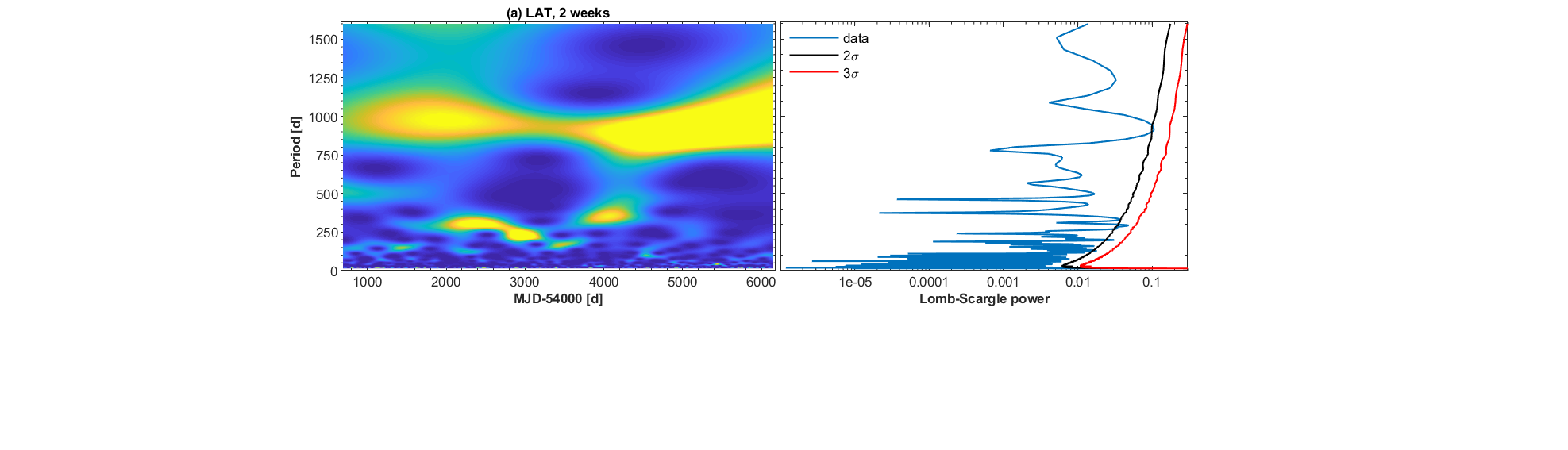}
 \includegraphics[trim=6.2cm 2.8cm 0.2cm 0cm, clip=true, scale=0.91]{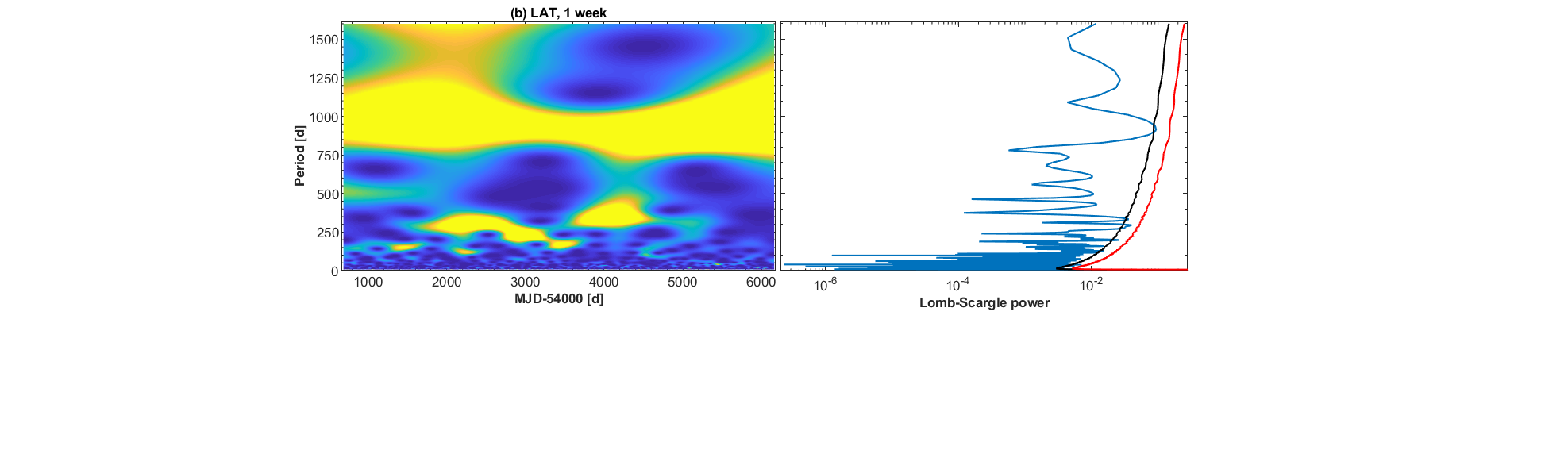}
 \includegraphics[trim=6.2cm 2.8cm 0.2cm 0cm, clip=true, scale=0.91]{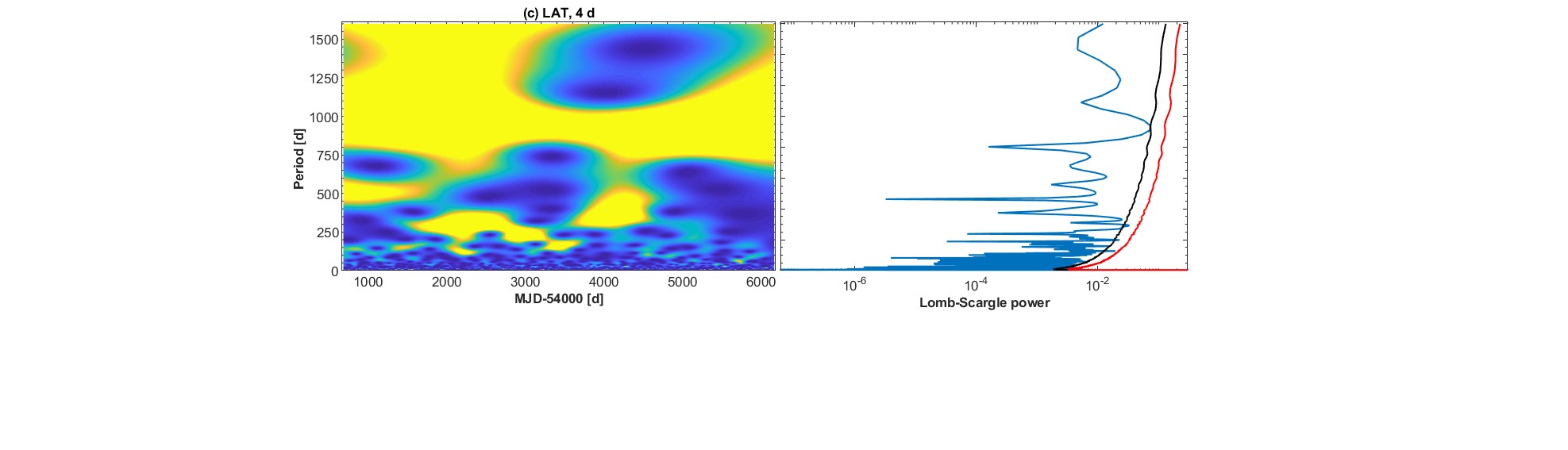}  \vspace{-0.9cm}
 \caption{\label{lsp} The WWZ (left column) and LSP (right column) plots from (1) the LAT observations during 2008--2023, based on the 0.3--300\,GeV flux values derived via the different time integrations.} \end{figure*}

Figure\,\ref{lsp} presents the WWZ and LSP plots of  corresponding to the LAT-band light curves of Mrk\,421 from the 2008--2023 period, constructed on the basis of different time integrations. In the WWZ plot, the possible period should emerge as a permanent narrow horizontal peak, corresponding to the most significant peak in the LSP plot (see, e.g., \citealt{o22,v18}). The latter also contains the curves corresponding to the detection significances with 2$\sigma$ and 3$\sigma$, determined by using the recipe of \cite{e13}. However, no highly significant periodical variability was found, since only the peaks with a significance of 3$\sigma$ and higher correspond to genuine periodicity \citep{o22}.  The significances of the existing PSD peaks, corresponding to some bright strips in the WWZ plots, do not reach the 3$\sigma$ significance.  The bright LSP stripe around $P$=1000\,d in Figure\,\ref{lsp}, changing in width  with time and shifting to larger values  after about MJD\,56500, can be explained as a red noise leak (see \citealt{o22}).

As for the past studies, \citep{bh19,bh20} claimed a  periodicity detection for Mrk 421 from the Fermi-LAT data. In addition to the issues related to the analysis of the LAT data (e.g., the energy range of 0.1–300 GeV instead of 0.3–300 GeV generally adopted for HBLs; see Section\,2.1), the reported periods (285 and 330 days) were detected below the 3$\sigma$ significance and/or showed some changes with time, as inherent to the red noise leak (similar to our result). Nevertheless, this detection was not confirmed by \cite{tar20} from the same 10-yr data by adopting some additional methods of the periodicity check.

The longer-term enhanced activity (e.g., those exhibited by Mrk\,421 in the LAT-band) could be result from the propagation and evolution of relativistic shocks through the observer-pointed jet (see, e.g., \citealt{bot19} ). In turn, a shock  can be triggered by the instabilities occurring in the innermost accretion disk, which  momentarily saturate the jet with the highly-energetic plasma carrying much larger pressure than the relatively steady jet plasma downstream (\citealt{s04} and references therein). This phenomenon could be reflected in a lognormal flaring activity of the source on various timescales, since  the latter may indicate a variability imprint of the accretion disk onto the jet (see, e.g., \citealt{rieg19}). Namely, independent density fluctuations can emerge in the disk on the local viscous timescales, characterized by a negligible damping. These instances can propagate in the direction of the innermost disc area,  merge there and produce a multiplicative behavior. The combined fluctuation can be transferred into  the jet flow (e.g., as an abrupt enhancement in the jet collimation rate), and the jet emission (including the LAT-band one) can be modulated correspondingly.  Consequently, a lognormal variability in the different energy range and over various timescales is then anticipated \citep{gieb09,rieg19}. 

The emission from the proton-induced synchrotron cascades (\citealt{m93} and references therein), or that from the magnetospheric IC pair production processes \citep{lev11} are also thought to yield log-normal flux distributions. However, there can be some limitations  by the gap travel time for the magnetospheric processes and from the dynamical or escape properties of the hadronic cascades  \citep{rieg19}. Furthermore, lognormal variability can be produced by random fluctuations in the particle acceleration rate \citep{sin18}. In that case, fluctuations in the acceleration rate can be also characterized by the Gaussian distribution of the photon-index along with the lognormal flux distribution. 

\begin{figure*}[ht!]
\includegraphics[trim=6.0cm -0.3cm -0.9cm 0cm, clip=true, scale=0.88]{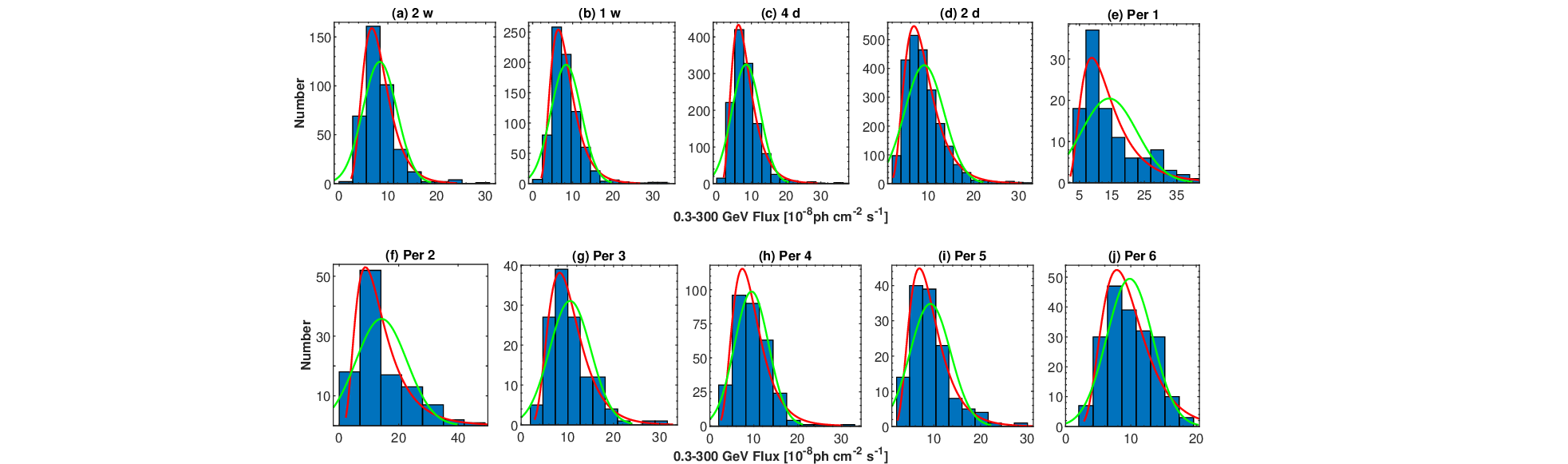}  
\includegraphics[trim=6.0cm 4.7cm -0.9cm 0cm, clip=true, scale=0.88]{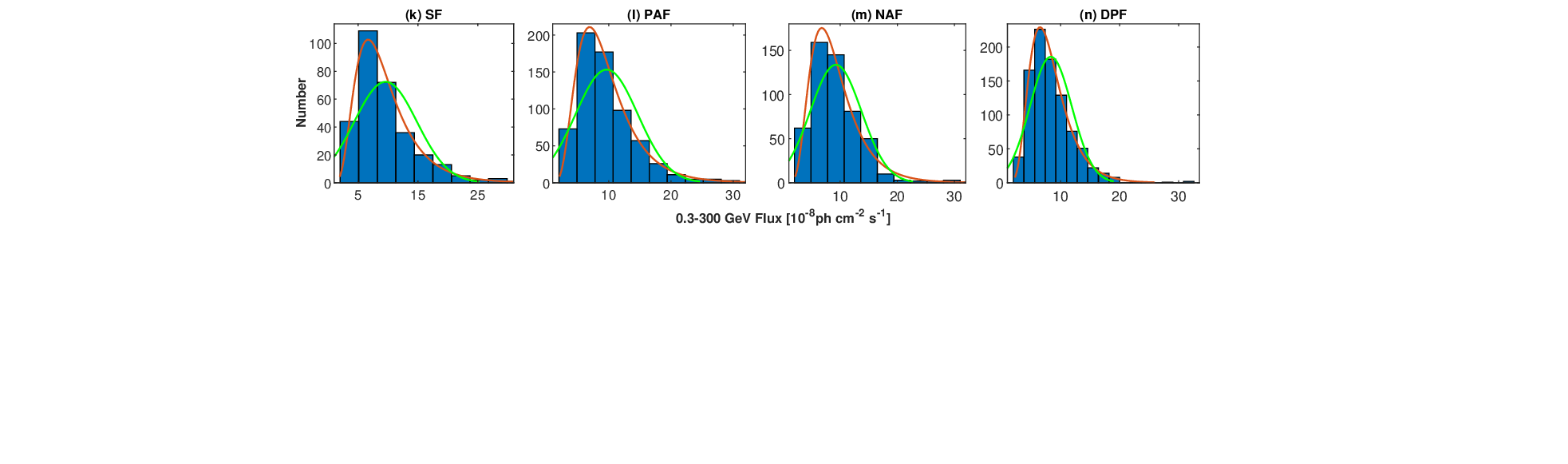}  
\vspace{-1cm}
 \caption{\label{logn} Distribution of the LAT-band flux values from the different time integrations during 2008--2023 [panels (a)--(d)]  and that  from the 2-d binned data for different periods [panels (e)--(j)], as well as for those time intervals showing the LAT-band flares of Mrk\,421 with different profiles [panels (k)--(n); SF -- symmetric flares, PAF -- positively-asymmetric flares, NAF  -- negatively-asymmetric flares and DPF -- double-peak flares]. The red and green curves show the lognormal and Gaussian fits to the histograms, respectively.}
 \end{figure*}

We checked the LAT-band light curves of Mrk\,421 constructed with different time integrations for the presence of lognormal variability.  Figures\,\ref{logn}a--\ref{logn}d and Table\,\ref{logntable} demonstrate that a lognormality was inherent to the target during the entire period of our study:  the lognormal function fits significantly better with the histograms constructed by using the integrations from two weeks down to one day than the Gaussian one. A similar situation was also for the samples  containing the 0.3--300\,GeV flux values from Periods 1--6, characterized by the different levels of LAT-band flaring activity of the target (Figures\,\ref{logn}e--\ref{logn}h). As noted in Section\,4, the distribution of the $\Gamma$-values sometimes was not very different from 
a Gaussian shape. Note that the lognormal flux variability along with a Gaussian distribution of the photon index can stem from the random fluctuations in the particle acceleration rate (see above). Note that a gradual (relatively slow) acceleration of the particles responsible for the IC upscattering of low-energy photons to the MeV--GeV range could be produced by stochastic (second-order) Fermi mechanism operating in the jet region with low magnetic field and high matter density \citep{vv05}. On the contrary, rapid injection of very energetic particles is inherent with the first-order Fermi mechanism within the Bohm’s limit of particle diffusion \citep{vv05}. We suggest that there could be frequent random transitions from dominance of the first-order Fermi process into that of stochastic acceleration and vice versa during the aforementioned periods. Note that the dominance of the Fermi-I process (operating in Bohm’s limit) is reflected in the clockwise (CW) spectral evolution of the flare in the flux--photon index plane \citep{tam09}. On the contrary, the source follows a counter-clockwise (CCW) spectral evolution when the Fermi-II process is dominant \citep{t09}. Note that the CW-to--CCW or converse transitions during single X-ray flares in Mrk\,421 in those periods were reported by \cite{k18b}. Similar  situations were evident also during other periods (see \citealt{k16,k17a,k18a,k20,k24}, detecting also the X-ray and FACT-band lognormality in the target). 

Note that the source sometimes showed symmetric short-term flares accompanied by very hard photon indices within some time bins, which could be powered by ”blobs” of magnetized, nonthermal plasma. In turn, such blobs can be produced by the RMR operating in the jet and not related to the AD instabilities (see below). The flux variability in the corresponding time intervals is not expected to show a lognormality and produce outliers from the histograms (as presented in  Figure\,\ref{logn}), along with those flux values containing the contribution from other local, purely jet-inherent processes. Namely, each histogram shows outliers or even low-amplitude secondary peaks at the fluxes higher than $\sim$1.5$\times$10$^{-7}$ph\,cm$^{-2}$s$^{-1}$. Moreover, the presence of those acceleration processes other than  those involving random fluctuations in the particle acceleration rate is reflected in deviations of the photon-index distributions from the Gaussian shape, as demonstrated by each histogram presented in Figures\,\ref{inddistr}--\ref{inddistr1}.

\begin{table*}[ht!] \centering     \tabcolsep 3.3pt    \begin{minipage}{170mm}
  \caption{\label{10gev} List of the  harder 10--300\,GeV spectra along with the corresponding MJD interval, $N_{\rm pr}$ and TS values.}     \begin{tabular}{cccccccccccc}        \hline
&\multicolumn{3}{c}{0.3--1\,GeV}&\multicolumn{3}{c}{1--10\,GeV} & \multicolumn{3}{c}{10--300\,GeV}  \\  
\hline
 Dates/MJDs  & $N_{\rm pred}$& TS  & $\Gamma$ &  $N_{\rm pred}$& TS  & $\Gamma$ & $N_{\rm pred}$& TS  & $\Gamma$ \\
 (1)	& (2) &	(3)&	(4) &	(5)&	(6)& (7)& (8)& (9)& (10)\\  \hline
2008\,Aug\,5—Sep\,1/(56)683--710&	108	&193	&1.57(0.15)&	86	&577	&1.66(0.09)&	14	&198	&1.37(0.12)\\
2008\,Sep\,30—Oct\,27/(567)39--66&	81	&133&	1.91(0.23)&	62&	319	&2.11(0.19)	&12	&143	&1.50(0.12)\\
2009\,Apr\,14—May\,11/(569)35--62&	85	&161	&1.89(0.20)	&63	&352	&1.94(0.16)	&16	&240&	1.61(0.10)\\
2009\,Jun\,9—Jul\,7/54991--55018&	114&	233&	2.16(0.18)&	82	&558	&1.99(0.15)	&13	&152&	1.72(0.12)\\
2009\,Aug\,4—31/(550)47--74	&100&	240&	1.53(0.14)	&48	&303	&1.85(0.17)	&11	&126	&1.54(0.12)\\
2010\,May\,11—Jun\,7/(553)27--54&	100&	196&	2.14(0.18)	&69	&463&	1.62(0.13)&	13	&197	&1.42(0.11)\\
2013\,Sep\,24—Oct\,21/(565)56--83&	114	&235	&1.72(0.18)	&79	&544	&1.64(0.12)&	16&	193	&1.37(0.12)\\
2014\,Jan\,14-Feb\,10/(566)71--88&	146&	285	&1.83(0.16)	&102	&589	&2.05(0.14)	&15	&194&	1.69(0.11)\\
2014\,Nov\,18—Dec\,15/56979--57006	&122	&348&	1.73(0.14)	&92	&713&	1.68(0.09)	&28	&388	&1.57(0.08)\\
2016\,Sep\,20—Oct\,17/(576)51--78&	42	&63	&1.95(0.26)	&13	&35	&2.72(0.38)	&8	&73&	1.65(0.16)\\
2018\,Jun\,26—Jul\,23/(58)295--322	&46	&74	&2.79(0.36)&	25	&131&	1.90(0.22)&	9	&121	&1.47(0.13)\\
2018\,Oct\,16—Nov\,12/(584)07--34	&88	&165&	1.77(0.20)	&70	&439	&1.87(0.17)	&10	&140	&1.34(0.14)\\
2018\,Dec\,11—2019\,Jan\,7/(584)63--90&	111	&194&	2.18(0.19)	&56	&256&	2.17(0.20)	&17&	242	&1.81(0.11)\\
2019\,Apr\,2—29/(58)575--602	&102	&165	&2.92(0.37)	&94&	606	&1.71(0.11)	&16	&264	&0.97(0.20)\\
2019\,Apr\,30—May\,27(586)03--30	&62	&100	&1.91(0.25)	&40	&178	&2.06(0.22)	&15&	253	&0.93(0.20)\\
2019\,May\,28—Jun\,24/(586)31--58)	&116	&220	&1.74(0.17)	&54	&311&	1.69(0.16)	&10	&124	&1.31(0.14)\\
2019\,Aug\,20—Sep\,16/(587)15--42	&39	&73	&2.62(0.36)	&59	&353	&1.94(0.16)&	8	&98	&1.18(0.18)\\
2019\,Nov\,12—Dec\,8/(58)799--826	&142&	327&	1.30(0.20)	&91	&553	&1.91(0.18)	&16	&202	&1.29(0.12)\\
2020\,Aug\,18—Sep\,14/(59)079--106	&60	&124	&1.83(0.24)&	33	&220&	1.70(0.18)&	8	&107	&1.41(0.14)\\
2021\,Sep\,14—Oct\,11/(594)71--98&	92	&185	&1.51(0.17)	&42	&215	&1.98(0.18)	&12&	130&	1.50(0.12)\\
2022\,Nov\,8—Dec\,5/(59)891--918	&163	&332	&1.94(0.15)&	130	&839	&1.87(0.11)&	24	&304&	1.47(0.10)\\
2022\,Dec\,6—2023\,Jan\,2/(599)19--46	&152&	372&	1.83(0.14)	&93	&674&	1.81(0.13)	&22	&305&	1.44(0.10)\\
2023\,Jan\,31—Feb\,27/59975--60002	&97	&237	&1.60(0.15)&	81&	541	&1.82(0.15)	&10	&155&	1.55(0.11)\\
 \hline \end{tabular} \end{minipage} \end{table*}

\begin{figure*}[ht!] 
\includegraphics[trim=6.0cm 0.1cm -1cm 0cm, clip=true, scale=0.9]{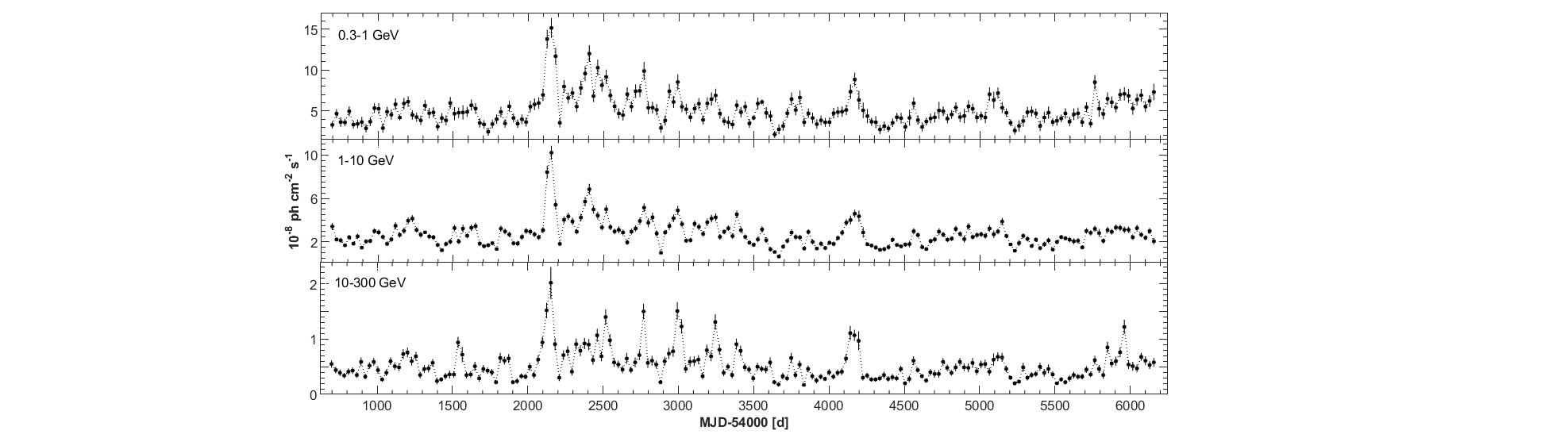}
\includegraphics[trim=6.0cm 4.3cm -1cm 0cm, clip=true, scale=0.9]{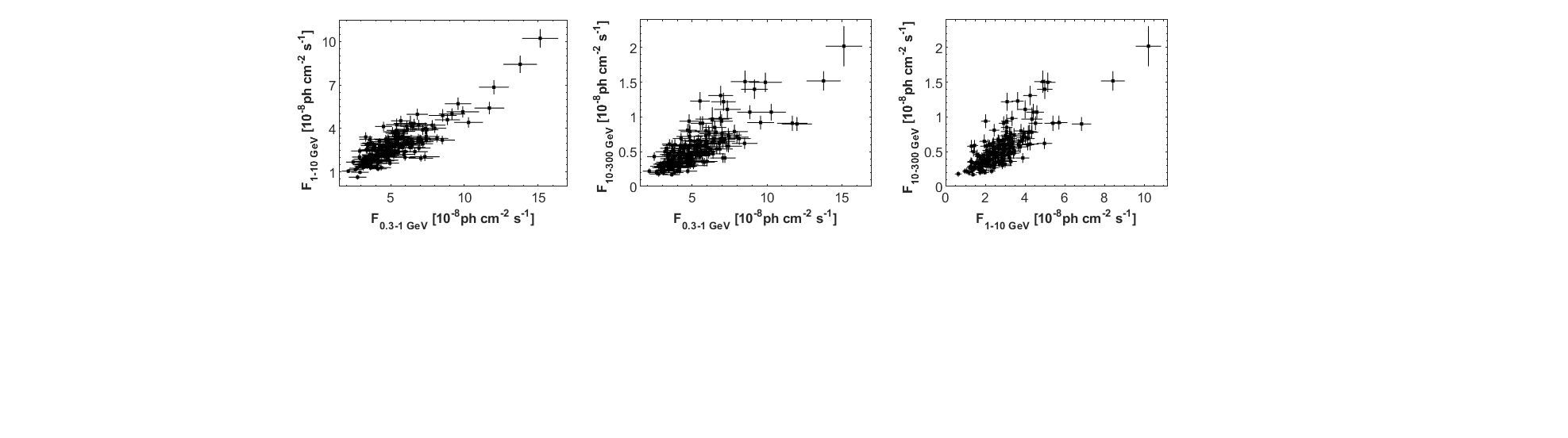} \vspace{-0.7cm}
 \caption{\label{subbands} Top: Light curves of Mrk\,421  in the separate 0.3--1\,GeV, 1--10\,GeV and 10--300\,GeV bands during 2008\,August--2023\,August, extracted with four-weekly integrations; bottom: correlations between the  0.3--1\,GeV, 1--10\,GeV and 10--300\,GeV fluxes. }
 \end{figure*}

As shown in Section\,3.1, short-term LAT-band flares were frequently seen in the epochs of the X-ray flaring activity, hinting at the connection between these instances, e.g., via the IC-upscatter of the X-ray photons in the Klein-Nishina (KN) regime. The later is notable for the suppression of the $\gamma$-ray emission and, consequently, the fractional variability amplitude was generally at least 50\% lower than that observed with XRT in the 0.3--10\,keV energy range (see Table\,\ref{periods}). Note that Table\,\ref{klein} presents  the time intervals of the 4-week duration when the LAT observations showed a softening at the energies beyond 1\,GeV or 10\,GeV compared to the lower-energy LAT-band part of the spectrum. These instances should be related to the IC-upscatter of  X-ray photons to the GeV energies in the KN-regime (versus the upscatter of lower-energy photons in the Thomson regime to the energies below 1\,GeV or 10\,GeV). Consequently, the light curves constructed for the separate 0.3--1\,GeV, 1--10\,GeV and 10--300\,GeV bands do not follow each other closely in the corresponding time intervals (Figure\,\ref{subbands}). Moreover, the possible KN-suppression weakened the expected strong correlation between the fluxes extracted from these bands down to $\rho$=0.63(0.07)--0.73(0.06): the data points corresponding to the instances provided in Table\,\ref{klein} produce outliers from the scatter plots $F_{\rm 0.3-1 GeV}-F_{\rm 1-10 GeV}$, $F_{\rm 0.3-1 GeV}-F_{\rm 10-300 GeV}$ and $F_{\rm 1-10 GeV}-F_{\rm 1-10 GeV}$ (see the bottom row of Figure\,\ref{subbands}).

A similar situation was also found with the MAXI-band variability in Periods\,2--4, hinting at the significant portion of the hard X-ray emission among the "seed" photons for the upscattering to the MeV--GeV energies in the KN-regime. Note also that the $F_{\rm var}$ values for the 1-10\,GeV and 10--300\,GeV bands from the intervals presented in Table\,\ref{klein} are lower than their "counterpart" in the 0.3-10\,GeV band, while the source generally showed a trend of higher $F_{\rm var}$ with increasing energy: $F_{\rm var}$=34.4(1.10) in the 0.3--10\,GeV band versus $F_{\rm var}$=50.1(1.1)\% beyond 10\,GeV for the entire set of the 4-weekly bins.

The exception was Period\,1 where the LAT-band $F_{\rm var}$ value was higher than its 0.3--10\,keV counterpart. However, no XRT observations were carried out in the epoch of the strongest 0.3--300\,GeV flaring activity (as discussed in Section\,3.1.1) and, consequently, no firm conclusion can be drawn. However, a similar situation (significantly lower fractional amplitude) was in the case of the regular MAXI observations: the 5$\sigma$-detections of Mrk\,421 were not frequent and occurred mostly during the  long-term LAT-band outburst, but the hard X-ray flaring activity was considerably weaker and the peak states were recorded only during the LAT-band decline epoch. Moreover, the regular BAT observations yielded only two 5$\sigma$-detections of the target in the epoch of the LAT-band outburst, showing relatively low 15--150\,KeV states. A significantly stronger BAT-band activity with more frequent 5$\sigma$-detections and much higher hard X-ray fluxes were recorded in other periods (except for Period\,1; see the corresponding panels in Figure\,\ref{per}). Similarly, higher MAXI-band states occurred in Periods 2, 6, 4, 8 and 10--11. Note that the strongest LAT-band outburst in 2012 coincided with onset of the strong, long-term \emph{V}-band flare. We suggest that the latter was probably triggered by the long-term, strong increase in the collimation rate of  leptons capable of producing a strong optical flare (but not an X-ray one). Consequently, an IC-upscatter of these lower-energy photons to MeV--GeV energies in the Thomson regime yielded a long-term boost in the LAT-band emission, which was not suffered of the KN-suppression due to the lack of corresponding ultrarelativistic electron population. 

Moreover, the MAXI-band fractional amplitude was  lower compared to the 0.3--300\,GeV counterpart also during Periods\,3--4 and 6, whereas a similar situation was observed also in the BAT band or the source did not show  variability at the 3$\sigma$ confidence level.  The difference was especially large in Period\,7: the MAXI-band $F_{\rm var}$ quantity was by a factor 2 lower than that derived from the LAT observations, and the flux variabilities recorded in these bands were obviously not correlated. Consequently, an IC-upscatter of hard X-ray photons to the MeV--GeV energies in the KN-regime  practically could not occur in that period.

Note also that the fluxes, corresponding to the highest LAT-band states in Periods\,1 and 3--5 produce outliers from the lognormal distribution. These states generally were recorded during the relatively fast flares superimposed on the long-term one. We suggest that these instances could be triggered by the shock interaction with the jet inhomogeneities, the origin of which was related to the jet instabilities (e.g., strong turbulent structures; \citealt{m14}). Since such structures have no relation with the AD instabilities, consequently, the associated highest LAT-band fluxes could not follow a lognormal distribution. 

\begin{table*}  \footnotesize  \begin{minipage}{170mm} 
  \caption{\label{hardening} Summary of the largest LAT-band hardening/softening instances, shown with arrows preceded with the maximum duration of the given instance (extract).}  \centering
  \begin{tabular}{ccccc}   \hline
 Dates/MJDs   & Photon index (error) /Maximum duration of the hardening/softening instance  \\     \hline
2008\,Aug\,5--15/(546)83--93	&1.65(0.11)[6d] $\rightarrow$ 2.31(0.22)[4d] $\rightarrow$ 1.33(0.15)[4d] $\rightarrow$ 2.32(0.22)[3d] $\rightarrow$ 1.60(0.13)\\ 
2008\,Aug\,19--27/(54)697--705&		1.39(0.12)[4d] $\rightarrow$ 2.26(0.21)[6d]$\rightarrow$ 1.33(0.13)\\
2008\,Sep\,14--30/(547)23--39	&	2.00(0.15)[10d] $\rightarrow$ 1.24(0.12)[4d] $\rightarrow$ 2.20(0.15)[6d] $\rightarrow$1.28(0.12)\\
2008\,Oct 7--22/(547)46--61&	1.62(0.13)[8d] $\rightarrow$ 2.26(0.16)[9d] $\rightarrow$ 1.54(0.13)\\
 \hline \end{tabular} \end{minipage} \end{table*}

Finally, the source showed a lognormal variability during the FACT observations in the different periods (see \citealt{k20,k24}). As discussed in Section\,3.1, Mrk\,421 frequently underwent simultaneous flaring activity in the VHE and 0.3--10\,keV energy ranges (and reported from a number of the MWL studies; see \citealt{h09,acc14,al15a,al15b,ah16} etc.) that is in accordance with with the one-zone SSC scenario and shows a generation of the corresponding emissions by the same electron population. However, some strong VHE flares in Periods 4--5 were not accompanied by the comparable 0.3--10\,keV "counterparts" (see Section\,3). The VHE flux peaked days before the X-ray one during the giant flare in 2004 that was impossible to explain via the standard one-zone SSC model, and \cite{bl05} suggested this instance to be an "orphan” TeV flare. \cite{acc11} also found the elevated X-ray states not being accompanied by TeV flaring and conversely  in 2006–2008 etc. These instances show that the one-zone SSC models was not always acceptable for the target. For example, a fast strong VHE flare without a significant simultaneous X-ray flaring activity on MJD\,57788 was  interpreted by \cite{acc21} as follows: the VHE  flare was caused by the appearance of a more compact second blob of highly-energetic electrons with considerably narrow energy range that could have been produced by stochastic acceleration, by the RMR, or by electron acceleration in the magnetospheric vacuum gap, close to the central SMBH. Note that the MAGIC energy range is partially overlapped by the 10--300\,GeV band used by us for constructing the light curve presented in the 3rd panel of Figure\,\ref{subbands}. However, the latter is based on the 4-weekly integration and, consequently,  the strong one-day VHE flare on MJD\,57788 is smoothed out, showing only a low-amplitude peak in that epoch. Note that the one-day binned data from the entire 0.3--300\,GeV band show the target's robust detection and a flaring state corresponding to (1.7$\pm$0.35)$\times$10$^{-7}$ph\,cm$^{-2}$s$^{-1}$ (not resolvable in the separate 0.3--1\,GeV, 1--10\,GeV and 10--300\,GeV sub-bands).

Note that the selection of these periods was based on relatively enhanced LAT-band flaring activity on timescales of several months to more than 1\,yr which, in turn, could be caused by an enhanced matter collimation rate on yearly timescales. This phenomenon could trigger also the baseline 0.3--10\,keV flux variability on yearly timescales  in our target (see \citealt{k24}), and it is even more clearly expressed in other nearby X-ray bright, TeV-detected HBLs Mrk\,501 and 1ES\,1959$+$650 \citep{k18c,k23}.

Another experimental confirmation for the shock presence in the jet of Mrk\,421 was provided by the 2--8\,keV observations with the Imaging X-ray Polarimetry Explorer (IXPE) carried out on 2022\,May\,4 \citep{d22}: the higher level of the X-ray linear polarization degree compared to longer wavelengths, and the absence of significant polarization variability was explained as a shock was the most likely X-ray emission site in the jet during that observation. 

Our study has revealed 175 instances of 0.3--300\,GeV flux doubling/halving   with the timescales $\tau_{\rm d,h}$ ranging from 5\,hr to $\sim$17.5\,d and generally associated with short-term flares. These instances were observed in all 6 periods discerned within our study (as well as in the intermediate time intervals), with significantly larger numbers in those periods characterized by the strong LAT-band activity of Mrk\,421. These detections allow us to constrain the upper limit to the size of LAT-band emission zone as (\citealt{sa13} and references therein)
\begin{equation}
 R_{\rm em}\leqslant {{c\tau_{\rm d,h} \Gamma_{\rm em}}/(1+z}) .   
\end{equation}
By assuming that the jet axis is aligned near the line-of-sight with the critical angle, then $\Gamma_{\rm em}$=$\delta$ \citep{b16} and we can adopt $\delta$=25 (the value of the Doppler factor frequently derived for Mrk\,421 from the different studies; see, e.g., \citealt{b16,acc21}) in Eq.\,(6). Consequently,  this yields a range of upper limits between  1.3$\times$10$^{16}$\,cm and 1.1$\times$10$^{18}$\,cm for the emission zone responsible for that extreme variability. Several of these instances included the 0.3--100\,GeV IDVs which occurred in Periods 1--2 and 5. Totally, we detected 25 IDVs from the LAT observations of Mrk\,421 with $F_{\rm var}$=36.7(10.5)--90.0(19.8)\,per cent from the robust detections of the source (with $TS$$\geq$9 and  $N_{\rm pred}$$\geq$8), and a vast majority of these instances  belongs to Periods 1--2 and 6. Note that only one out of these IDVs (occurring in 2013\,April) was reported within the past  studies (see \citealt{k16}), and Mrk\,421 is the only HBL source which has shown a LAT-band IDV (owing to the general presence of the higher-energy SED peak beyond the LAT range, contrary to the LBL and IBL objects). The highest-amplitude 0.3--300\,GeV IDV was associated with the brightness drop by a factor more than 3.2 (taking the error ranges into account; on  2022\,May\,8, Period\,6), whereas the fastest instance was recorded on 2013\,July\,15 (during the unprecedented X-ray and TeV-band outbursts): after reaching the highest historical level of 10$^{-7}$ph\,cm$^{-2}$s$^{-1}$, the 0.3--300\,GeV brightness practically halved within the next 2\,hr. However, these instance were significantly less extreme compared to those recorded in the X-ray and TeV energy ranges (generally containing the lower- and higher-energy SED peaks of Mrk\,421). For example, the VHE flux increased by a factor of 20–25 in about 0.5\,hr \citep{g96}. The source varied within the time intervals as short as a few hundred second in the 0.3--10\,keV band and showed a flux doubling/halving instances down to  1\,hr in 2013\,April \citep{k16} etc.

\subsection{Particle acceleration and cooling}

As discussed in Section 3.1, the 0.3--300\,GeV light curves were characterized by a variety of flare profiles (symmetric, two-peak, positive or negative asymmetry). Different studies showed that  the specific cases of the interplay between the  particle acceleration and cooling  can yield a characteristic profile of the particular flare. First of all, a symmetric profile can be determined by the light travel time effects, while the particle acceleration and cooling timescales (in the given spectral range) are much shorter than the light-crossing timescale \citep{f24}. This means that these timescales were much shorter than the light-crossing time the LAT-band emission zone for the (quasi)symmetric flares presented in Table\,\ref{shortterm}. 

Alternatively,  a symmetric shape of the flare (with a possible plateau) indicates that the observed variability was driven by the crossing time-scale of the underlying disturbance, e.g., a shock front  \citep{r19}. Moreover,  relativistic magnetic reconnection in blazar jets can produce “plasmoids”, i.e., "blobs" of magnetized, nonthermal plasma of various sizes, which contain high-energy  particles capable of upscattering lower-energy photons to the $\gamma$-ray energy range  (see, e.g., \citealt{p15,p16}). According to time-dependent modeling of emission from these extreme jet structures, the plasmoid-powered $\gamma$-ray flares (e.g., in the LAT energy range) can be observed as symmetric in the case the plasmoids are not changing in size rapidly \citep{chr19,may21}. It is important that a hard or very hard power-law electron energy distribution (EED) $N(\gamma$$\propto\gamma^{-p}$ can be established by the relativistic reconnection operating in the magnetized  jets areas, characterized by $p\rightarrow$1 when the photon index $\Gamma$$\lesssim$1.5 and the upstream magnetization   $\sigma_{\rm up}$$\gtrsim$10  (see, e.g., \citealt{sir14}).

\begin{table*}[ht!]  \tabcolsep 2.7pt  \begin{minipage}{170mm}
  \caption{\label{klein} List of the time intervals characterized by spectral softening at the GeV energies (extract). Columns 2--4 present the number of the model-predicted photons, test-statistics and photon-index value in the 0.3--1\,GeV band, whereas the same quantities from the 1--10\,GeV and  10--300\,GeV bands are provided in the Columns 5--7 and 8--10, respectively.}   \centering
   \begin{tabular}{cccccccccccc}        \hline
&\multicolumn{3}{c}{0.3--1\,GeV}&\multicolumn{3}{c}{1--10\,GeV} & \multicolumn{3}{c}{10--300\,GeV}  \\  
\hline
 Dates/MJDs  & $N_{\rm pred}$& TS  & $\Gamma$ &  $N_{\rm pred}$& TS  & $\Gamma$ & $N_{\rm pred}$& TS  & $\Gamma$ \\
 (1)	& (2) &	(3)&	(4) &	(5)&	(6)& (7)& (8)& (9)& (10)\\
 \hline
2009\,May\,12—Jun\,8/(549)63--90&	157	&342	&1.53(0.10)&	114&	705	&1.93(0.12)&	23&	303&	1.78(0.10)\\
2009\,Jul\,7—Aug\,4/(550)19--46&	82	&124	&1.59(0.19)&	91	&578&	1.57(0.09)&	10&	174	&2.68(0.20)\\
2009\,Dec\,22—2010\,Jan\,18/(55)187--214&	145	&372&	1.36(0.15)&	122	&879	&1.43(0.08)&	24	&258	&1.69(0.10)\\
2010\,Jan\,19—Feb\,15/(552)15-42&	88&	229	&1.27(0.20)&	121	&888	&1.53(0.05)&	15&	195	&1.71(0.11)\\
  \hline \end{tabular} \end{minipage} \end{table*}

 We have checked that the source showed very and extremely hard LAT-band spectra during some symmetric flares. This especially was the case during the shorter-term instances lasting several days [e.g., those occurring during MJD (548)46–52, (553)25–31 and  (572)33—40] when only very and extremely hard LAT-band spectra were observed. These flares could be produced by those plasmoids containing very energetic plasma and characterized by slow change in size. Note that \cite{k24} presented a number of the detections of the reconnection-related features from the X-ray spectral study of mrk\,421, which belong to each period discussed in Section\,3.2. However, other instances also include   softer spectra  or are exceedingly long to be triggered only by the RMR, and could be affected by the light time travel effects. Finally, logparabolic spectra were also observed during some symmetric flares. Such spectra can be established within the energy-dependent acceleration probability process (EDAP), which represents a variety of the first-order Fermi mechanism operating at the relativistic shock front (\citealt{m04} and references therein):  electrons can be confined by a magnetic field at the shock front, while the confinement efficiency is dropping with increasing lepton's energy (Otherwise, the establishment of power-law EEDs are expected). Moreover, the logparabolic EEDs can be established by the stochastic acceleration, which can be efficient in the jet area with strong magnetic turbulence. The latter can be strongly enhanced after the passage of a relativistic shock in the magnetised, inhomogeneous jet medium \citep{m14}. Different observational features, demonstrating the importance of  both processes,  have been reported by various authors from intense X-ray spectral studies, covering most of the periods discussed in Section\,3.1 (e.g., \citealt{t09,k16,k17a,k18a,k18b,k20,k24}). Therefore, some symmetric flares could be driven by the time-scale required for a relativistic shock front to cross that jet area, which was characterized by those physical properties required for electron acceleration to the energies sufficient to upscatter low-energy photons to the MeV--GeV energies.  
  
Moreover, a symmetric flare profile can be the result of the superposition of several episodes of short duration \citep{ab10b}, and each one may be related to the different aforementioned processes. Finally, extremely hard LAT-band spectra could be the result of the significant contribution from the photons produced in the framework of the different hadronic processes (see the discussion below). Among the different periods, the symmetric LAT-band flares were the most frequent in Period\,6 (four flares), leading to the suggestion that they were relatively favourable in point of the physical conditions yielding such instances. On the contrary, Period\,5 was notable for the least occurrence of symmetric flares (only one instance in each). Note that the 0.3--300\,GeV flux showed a lognormal variability during  the symmetric flares, and the distribution of the corresponding photon-index values does not show a large difference from the Gaussian distribution (see Figures \ref{inddistr1}g and \ref{logn}k). This result leads to the suggestion that these flares were predominantly controlled by the shock crossing time-scales and affected significantly by random fluctuations in the  particle acceleration rate. These fluctuations could be caused by the subsequent shock passage through the jet area with different magnetic field properties (as discussed above). On the contrary, such distributions are lesser expected within the the plasmoid-related processes triggered by the local, pure jet-related instabilities.

Along with absence of the suitable physical conditions, the observation of symmetric flares is limited in the framework of the multi-zone emission scenarios: in the case the different emission regions are situated at different azimuthal angles in the jet cross-section and even each produce a symmetric flare, one could observe an overall asymmetric variability profile after the  superposition of these emissions  \citep{n13}. Namely, the time-dependent modeling of \cite{s15} demonstrated that a significant non-uniformity of the Doppler factor across the jet (caused by the radial expansion of the flow at the emission zone and boosted by the relativistic effects) can yield a significant symmetry distortion of the observed light curves and produce an positively-asymmetric flare profile with substantially extended brightness declining phase: since the emitting "shells" are considerably extended in the jet radial direction, their different parts are observed at different viewing angles. Consequently, the emission produced within those parts characterized by the largest inclinations arrive to the observer with a significant time delay compared to the emission situated at smaller viewing angles. In such a situation, the light  curve corresponding to this phase can be  less impacted by the radiative cooling of the highest-energy particles and  will be dependent on the gradient of the bulk Doppler factor across the emitting shells. We suggest that this effect could be especially important during the periods notable for a lack of symmetric flares.

However,  the variance of the Doppler factor across the emitting shell becomes smaller with the decreasing jet opening angle. Consequently, the flare asymmetry also decreases and is expected to become symmetric for the jet opening angle $\theta \lesssim$0.3\,deg. Although \cite{w22} obtained a large value of the jet opening angle from the VLBA observations of Mrk\,421 at 43\,GHz ($\sim$55\,deg), this result was mostly due to the projection effects caused by the very small angle to our line-of-sight (estimated to be $\sim$1.4\,deg). The de-projected value, based on the method of \cite{j17}, yields $\theta$$\sim$1.4\,deg  which is lower than the aforementioned threshold. Nevertheless, the jet of  Mrk\,421 is expected to be significantly narrower at the location of the $\gamma$-ray emission zone (situated much closer to the central SMBH than those jet parts emitting at 43\,GHz). Moreover, the jet width at this location is possibly variable from epoch to epoch, becoming narrower than the aforementioned threshold and allowing us to observe symmetric LAT-band flares during some time intervals. 

Alternatively, the origin of positively-asymmetric flares can be related to a fast injection of accelerated particles and slower radiative cooling and/or escape from the energization region. Consequently, such instances should be governed by Fermi-I acceleration at the relativistic shock front. Note that the distribution the 0.3--300\,GeV flux values and photon-index from the corresponding time intervals are in favour of this possibility (see Figures \ref{inddistr1}h and \ref{logn}l). On the other hand, the radiative lifetimes corresponding to the MeV--GeV energy range are generally very short. When the emitting shells are not considerably extended in the jet radial direction during the particular flare, the observed flare profile with an apparent positive asymmetry can by produced by the superposition of two or more  low-amplitude, shorter-term events  occurring during the long-term brightness decline \citep{r19}. Note that the LAT-band flares with a positive asymmetry were relatively numerous in Period\,5, characterized by a lack of symmetric flares. The latter results could be related to the superposition of lower-amplitude symmetric flares, producing a single, brightness-declining profile with a significantly longer duration compared to the brightness-increase phase.  For example, strongly asymmetric flares frequently showed a subsequent secondary maximum after the peak brightness (see Figure\,\ref{sym}B). According to \cite{n13}, the emitting region situated at the same distance across the jet, but oriented at a smaller angle to our line-of-sight, is located closer to the observer. Consequently, the light travel time is shorter for the emission from this region and is more strongly
Doppler boosted compared to that from the region situated at larger distance from the jet axis. Therefore, one can observe a major peak followed by the minor one. Note that this scenario also can yield the distribution presented in  Figure\,\ref{logn}l, since the superposing symmetric flares could be governed by the shock-crossing timescale and show a lognormal flux distribution (as discussed above).

On the contrary, flares with a negative asymmetry may indicate a gradual acceleration of the electrons  responsible for the IC upscattering of low-energy photons to the MeV--GeV range: the cooling time-scale of these particles can be shorter than the acceleration one \citep{r19}. As mentioned above, a gradual electron acceleration  is found to be inherent to the stochastic mechanism operating in the jet region with a low magnetic field and high matter density. Note that the negatively-asymmetric flares occurred in all periods, hinting at the  importance of the stochastic particle acceleration [as reported by \cite{k16,k17a,k18a,k18b,k20,k24} from the intense X-ray spectral study]. On the other hand,  the observation of longer rising phase of the flares can be simply due to the superposition of two or more  low-amplitude and short symmetric instances, not individually detectable but producing an apparently prolonged rising phase of a  single flare (see \citealt{r19}). Note that the flux and photon-index distributions from the time intervals of negatively-asymmetric flares (Figures \ref{inddistr1}i and \ref{logn}m) hint at the viability of both these scenarios (strong turbulence due in the shocked jet area and $\gamma$-ray flare controlled by shock-crossing timescale).

Finally,  the source also  exhibited  a two-peak  flare profile 36 times in the LAT-band, which were observed during all here-discussed periods. According to the semi-analytic  internal-shock model of \cite{b10}, two-peak flares can be related to the propagation of forward and reverse shocks. Namely, the central engine is considered to eject intermittently  "shells" of high-energy, relativistic plasma at varying speeds through the blazar jet, which subsequently collide. Consequently, two different shocks may appear: a forward shock moving into the slower shell and a reverse one propagating in the faster shell. According to these simulations, (i) the higher-energy end synchrotron peaks, established directly behind the forward and reverse-shock fronts, remains essentially unaffected as long as the observer receives synchrotron emission from the shocks still being located within the shells; (ii) as the forward and reverse shocks propagate, an increasingly larger region of the shells is energized with particles having longer time to cool. Consequently, the synchrotron spectrum extends progressively from X-ray to UV and lower frequencies; (iii) as one observes the shock regions leaving the shells, the highest-energy electrons rapidly cool and leading to the decline in the high-frequency synchrotron emission. One expects a delayed response of the SSC component with respect to the X-ray emission, with slightly cooled electrons still being able to efficiently upscatter synchrotron "seed" photons up to $\gamma$-ray energies in the Thomson regime. Note that this scenario frequently was the case for the  two-peak  flares presented in Table\,\ref{shortterm} (see also the MWL light curves from different periods in Figure\,\ref{per}).

Since the emergence of colliding plasma shells could be caused by those unstable processes which operate in the innermost AD parts, the resulted double-peak flares should show their imprint on the target's jet. Actually, the distribution of the corresponding 0.3--300\,GeV flux values clearly prefer the lognormal function (see  Figure\,\ref{logn}n and Table\,\ref{distrtable}). Moreover, some double-peak flares could be simply a superposition of two separate LAT-band flares, the origin of which were not related to the colliding shells but to some processes capable for yielding a lognormal variability (as discussed above). For example, a subsequent passage of single relativistic shock (triggered by the AD instabilities) through those jet inhomogeneities, which were  characterized by different sizes and magnetic field properties but situated relatively closely to each other. Consequently, two separate  flares with similar or different profiles (symmetric, positive  or negative asymmetry) could occur, which overlapped each other and observed as a single double-peak flare. Eventually, the corresponding fluxes would also follow the lognormal distribution (as presented in  Figure\,\ref{logn}n). The distribution of the photon-index from the epochs of double-peak flares is different from the Gaussian function, to be observed within the dominance of a single process (random fluctuations in the particle acceleration rate; see Figure\,\ref{inddistr1}j).

\subsection{Origin of the hard LAT-band spectra}

As noted above, very and  extremely hard $\gamma$-ray spectra are most commonly explained to have a  hadronic origin (see, e.g., \citealt{m93,sh15}). Accelerated leptons and hadrons are injected in the emitting region, which is uniformly filled with a magnetic field of strength $B$.  These assumptions are in accord with the one-zone SPB model: all radiation mechanisms are operating in the same emission zone and external photon fields are negligible (see, e.g., \citealt{c20}). The proton-proton interactions are thought to be negligible in the SPB models, since this mechanism requires very high particle density and the extreme jet powers for producing a significant $\gamma$-ray emission \citep{sol22}.  Table\,\ref{10gev} presents the time intervals characterized by harder spectra in the 10--300\,GeV energy range and explained by the significant contribution from the photons emitted in the framework of the proton-induced hadronic cascades (generally characterized by timing/spectral variability on longer timescales; see, e.g., \citealt{sh15,sol22}).  

On the other hand, these instances can be explained within the framework of modified the one-zone SSC model of  \cite{z21}, which assumes that electrons are co-accelerated with protons by relativistic recollimation shocks under physical situations as follows: (1) low jet magnetization and (2) electrons can be preheated in the shock transition layer, yielding relatively large minimum Lorentz factors when involved in the Fermi-I process.  The latter can produce high-energy electron populations characterized by a large range of power-law indices down to very hard ones ($p\simeq$1), depending on the properties of magnetic field and turbulence, shock speed and  obliquity \citep{sum12}. Namely, oblique, relativistic shocks (referred to as "superluminal", implying that they cannot be the sites via the mutual Fermi-I mechanism) can energize charged particles via shock-drift acceleration (SDA; \citealt{beg90} and references therein), which is also known as  fast Fermi process: particles are allowed to boost their energy by an order of magnitude even during a single shock encounter \citep{sol22}. When the MHD turbulence is relatively weak, the SDA is the most efficient and can produce very or extremely hard EEDs (see, e.g., \citealt{sir13}). However, the $\gamma$-ray emission from the accelerated proton population (with the same number density as the electrons) should not make a significant contribution in the total energy "budget".

Note also that the photon–photon absorption process can yield arbitrarily hard spectra, if the $\gamma$-ray emission passes through the medium containing a hot photon gas with a narrow energy distribution characterized by $E_{\gamma}\epsilon_0\gg m_{\rm e}c^2$ \citep{ah08}: the medium becomes optically thick at the lower $\gamma$-ray energies and thin at higher one (due to the decrease in the cross-section of the $\gamma\gamma$ interaction). Consequently,  intrinsically hard $\gamma$-ray spectra can be established. Moreover, \cite{l11} presented a time-dependent SSC model where extremely hard electron distribution  is achieved within the stochastic acceleration of electrons producing a steady-state, relativistic  Maxwellian (RM) particle distribution.  The latter represents a time-dependent solution of the Fokker-Plank equation that incorporates the adiabatic and radiative energy losses of accelerating particles. Depending on the physical conditions in the jet emission zone (e.g., if particles undergo cooling beyond the acceleration zone, or  the jet medium is clumpy), the combination of different pile-up distributions are capable of interpreting the observed $\gamma$-ray spectra. \cite{sh16} presented a two-zone SSC scenario for the very and extremely hard LAT-band spectra of the HBL source Mrk\,501, incorporating (1) a slowly-variable shock-in-jet component producing the underlying softer spectrum  through the Fermi-I first process and (2) fastly variable, very hard component produced by  intermittent injection of sharply peaked RM-type EED originating from the jet base. The latter could be established through stochastic acceleration from randomly moving Alfvén waves. This physical situation  could be sometimes the case for also Mrk\,421 when exhibiting very and extremely hard gamma-ray spectra. Moreover, very and extremely hard power-law spectra can be established by the RMR. Note that the importance of stochastic electron acceleration and RMR in the jet of Mrk\,421 is reported in our previous studies and discussed in Sections\,5.1--5.2. 

The source showed large changes of the spectral hardness many times, with the most extreme hardenings and subsequent softenings with $\Delta \Gamma$$>$1 during a few days. These instances (as well as those observed on longer timescales) were generally associated with the emergence of the extremely hard spectra. The latter could have been established by the turbulence-driven RMR, while the turbulence had been strongly enhance by the shock passage through the magnetized jet medium. On the other hand, small-scale strong turbulence could produce the spectral transitions power-law $\rightarrow$ logparabolic $\rightarrow$ power-law, which were shown by  the source many times during the 15-yr period.  Some other fast hardenings/softenings could be established by the subsequent emergence of the plasmoids corresponding to the relativistic and non-relativistic (yielding softer power-law spectra; \citealt{f23}) magnetic reconnections in the $\gamma$-ray emission zone.

\section{Summary} 
In this paper, we have presented the results from the detailed timing and spectral study of the nearby HBL source Mrk\,421, based on the data obtained from the practically uninterrupted 15-yr \emph{Fermi}-LAT observations during 2008\,August--2023\,August. The  experimental results are compared with those obtained within different theoretical studies that made us capable of drawing conclusions about unstable physical processes operating in the jet emission zone and possible hadronic contribution to the observed HE emission. For this purpose, we also examined the interplay between the LAT-band flux variability and those observed in the VHE $\gamma$-rays (the data obtained with FACT and various Cherenkov-type telescopes), X-rays (XRT, BAT and MAXI), optical--UV (UVOT and ground-based telescopes) and radio (UMRAO, VLBA) energy ranges.  Our basic experimental results and the underlying plausible physical processes can be summarized as follows:

\begin{itemize} \vspace{-0.2cm}
\item During the 15-yr period, Mrk\,421 was the brightest LAT-band source among HBLs and detectable down to intraday timescales during the strong HE flares. The mean 0.3--300\,GeV photon flux was about 7$\times$10$^{-8}$ph\,cm$^{-2}$s$^{-1}$, attained the maximum value of $\sim$10$^{-6}$ph\,cm$^{-2}$s$^{-1}$ during the two subsequent 1-hr segments during the 2013\,April outburst and frequently was brighter than 10$^{-7}$ph\,cm$^{-2}$s$^{-1}$ (observed very rarely for other HBLs).

\item The source showed non-periodical brightness changes on various timescales. The longest-term variability was related to the baseline 0.3--300\,GeV brightness change  from several to yearly timescales, superimposed by shorter-term flares lasting from about one week  to almost 4\,months. Depending on the baseline flux variability and the strength of flaring, we discerned six periods of the stronger 0.3--300\,GeV flaring activity. The variable baseline brightness was superimposed by 32 symmetric and 37 two-peak flares (during the entire 2008--2023 period), as well as  by those having positive and negative asymmetries (42 and 37 instances, respectively), characterized by fractional amplitudes $F_{\rm var}$=15.5(5.1)--72.7(7.8)\,per cent. The strongest MeV--GeV flaring activity of Mrk\,421 was recorded during 2012\,June--2013\,October and 2017\,October--2018\,March. Our study  revealed 175 instances of 0.3--10\,GeV flux doubling/halving   with the corresponding timescales $\tau_{\rm d,h}$ ranging from 5\,hr to $\sim$17.5\,d and generally associated with short-term flares. These instances allowed  us to constrain the upper limit to the  emission zone size as 1.3$\times$10$^{16}$\,cm--1.1$\times$10$^{18}$\,cm by assuming a Doppler factor of 25. Several out of these instances included the 0.3--10\,GeV IDVs, the total number of which amounted to 25  and were characterized by $F_{\rm var}$=36.7(10.5)--90.0(19.8)\,per cent.

\item Disparate flare profiles are related to the different interplays between the light-crossing, particle acceleration and cooling timescales. Symmetric flares are governed by the light time travel effects or by the crossing time-scale of the underlying disturbance, as well as could be produced by the reconnection-born plasmoids not changing in size rapidly. Relatively long symmetric flares can be the result of the superposition of several episodes of short duration.  However, when the different emission regions are situated at different azimuthal angles in the jet cross-section and even each produce a symmetric flare,  one can observe a positively-asymmetric flare profile. Alternatively, the origin of positively-asymmetric flares can be related to a fast injection of accelerated particles and slower radiative cooling and/or escape from the energization region. On the contrary, flares with a negative asymmetry may indicate a gradual acceleration of the electrons  responsible for the IC upscattering of low-energy photons to the MeV--GeV range: the cooling time-scale of these particles can be shorter than the acceleration one, hinting at the  importance of the stochastic particle acceleration. The source also  frequently exhibited   two-peak  flare profiles, plausibly  triggered by the "shells" of high-energy, relativistic plasma moving with different speeds through the blazar jet and subsequently colliding. 

\item During the period of our study, the source underwent a lognormal variability in the LAT energy range, explained as an imprinting of the disc nonstationary processes on the jet (e.g. shock propagation through the jet,  triggered by the multiplicative processes in the disc innermost parts).  Alternatively, the lognormal variability could be contributed also by proton-initiated hadronic cascades, the presence of which explains the observation of very hard LAT-band spectra and hardening beyond 10\,GeV.  Moreover, a lognormal flux variability was accompanied by the Gaussian distribution of the photon index during some periods, explained by random fluctuations in the particle acceleration rate (in turn, possibly due to  the frequent random transitions from dominance of the first-order Fermi process to that of stochastic acceleration and vice versa). Short-term LAT-band flares were frequently observed in the epochs of the X-ray flaring activity, hinting at the connection between these instances, e.g., via the IC-upscatter of the X-ray photons in the KN-regime. The latter explains a spectral softening at the energies beyond 1\,GeV or 10\,GeV compared to the lower-energy part of the spectrum during some time intervals and attenuation of the expected strong correlation between the fluxes extracted in these bands. 

\item Most of the 0.3--300\,GeV spectra were well-fit with a simple power-law model and showed a very broad range of the photon-index from $\Gamma$$\sim$2.8 down to $\Gamma$$\sim$1.2, with the mean values $\Gamma_{\rm mean}$=1.75(0.01)--1.84(0.01) and distribution peaks $\Gamma_{\rm p}$=1.73(0.01)--1.82(0.01).
The source showed large changes of the spectral hardness changes, with the most extreme hardenings and/or softenings with $\Delta \Gamma$$>$1 during 2-5 days, generally associated with the emergence of the extremely hard spectra. The latter could be established by the turbulence-driven RMR, by the SDA at recollimation shocks or  by the subsequent emergence of the plasmoids corresponding to the relativistic and non-relativistic magnetic reconnections in the $\gamma$-ray emission zone. Strong small-scale turbullence could trigger the spectral transitions power-law $\rightarrow$ logparabolic $\rightarrow$ power-law, as well as produce very logparabolic spectra with a wide range the photon index down to $\alpha$$\sim$1. The hardenings, observed on relatively longer timescales during those time intervals characterized by harder spectra in the 10--300\,GeV energy range could be associated to the proton-induced hadronic cascades. 
\end{itemize}
 
\section{Acknowledgements} 
BK and AG thank Shota Rustaveli National Science Foundation and E. Kharadze National Astrophysical Observatory (Abastumani, Georgia) for the fundamental research grant FR$-$21--307.  We acknowledge the use of public data from the \emph{Swift} data archive. This research has made use of the \texttt{XRTDAS} software, developed under the responsibility of the ASDC, Italy.  We acknowledge the use of the VHE data from long-term Whipple observations and the variable star observations from the AAVSO International Database contributed by observers worldwide and used in this research. The observational program at UMRAO was  supported in part by a series of grants from the NSF, most recently AST 0607523, and by a series of grants from the NASA Fermi Guest Investigator program NNX 09AU16G, NNX10AP16G, NNX11AO13G, and NNX13AP18G. This study makes use of VLBA data from the VLBA-BU Blazar Monitoring Program (BEAM-ME and VLBA-BU-BLAZAR;
http://www.bu.edu/blazars/BEAM-ME.html), funded by NASA through the Fermi Guest Investigator Program. The VLBA is an instrument of the National Radio Astronomy Observatory. The National Radio Astronomy Observatory is a facility of the National Science Foundation operated by Associated Universities, Inc.  Finally, we thank the anonymous referee for his/her useful suggestions.

\end{document}